\RequirePackage{fix-cm}
\documentclass[smallextended]{svjour3}       % onecolumn (second format)
\smartqed  % flush right qed marks, e.g. at end of proof
\usepackage{graphicx}
\usepackage{enumerate}
\usepackage{amsmath}
\usepackage{subfigure}
\usepackage{theorem}
\usepackage{multirow}
\newcommand{\tabincell}[2]{\begin{tabular}{@{}#1@{}}#2\end{tabular}}
\usepackage{algorithm}
\usepackage{algorithmic}

\begin{document}

\title{A Novel Quantum Image Compression Method Based on JPEG
\thanks{This work is supported by the National Natural Science Foundation of China under Grants No. 61502016 and 61602019, and the Fundamental Research Funds for the Central Universities
under Grants No. 2015JBM027.}}


\author{Jian Wang \and Nan Jiang \and Na Zhao}


\institute{J. Wang \at
              School of Computer and Information Technology, Beijing Jiaotong University, Beijing 100044, China \\
              Department of Computer Science, Purdue University, West Lafayette
              47907, USA \\
              Department of Chemistry, Purdue University, West Lafayette
              47907, USA \\
              \\
           N. Jiang \at
              College of Computer Science, Beijing University of Technology, Beijing 100124, China \\
              Department of Computer Science, Purdue University, West Lafayette
              47907, USA \\
              Department of Chemistry, Purdue University, West Lafayette
              47907, USA \\
              Beijing Key Laboratory of Trusted Computing, Beijing 100124, China \\
              National Engineering Laboratory for Critical Technologies of
Information Security Classified Protection, Beijing 100124, China \\
              \email{jiangnan@bjut.edu.cn}\\
              \\
           N. Zhao \at
              College of Computer Science, Beijing University of Technology, Beijing 100124, China
}

\date{Received: date / Accepted: date}

\maketitle

\begin{abstract}
Quantum image processing has been a hot topic. The first step of it
is to store an image into qubits, which is called quantum image
preparation. Different quantum image representations may have
different preparation methods. In this paper, we use GQIR (the
generalized quantum image representation) to represent an image, and
try to decrease the operations used in preparation, which is also
known as quantum image compression. Our compression scheme is based
on JPEG (named from its inventor: the Joint Photographic Experts
Group) --- the most widely used method for still image compression
in classical computers. We input the quantized JPEG coefficients
into qubits and then convert them into pixel values. Theoretical
analysis and experimental results show that the compression ratio of
our scheme is obviously higher than that of the previous compression
method. 
\end{abstract}

\section{Introduction}
Recently, quantum image processing has attracted a lot of attention.
In general, it has three steps as Fig. 1 shows: 1) store the image
into a quantum system, which is also known as preparation; 2)
process the quantum image; and 3) obtain the result by measurement.
In this paper, we will focus on quantum image preparation and focus
on increasing preparation efficiency.

\begin{figure}[h]
  \centering
  \includegraphics[width=8cm,keepaspectratio]{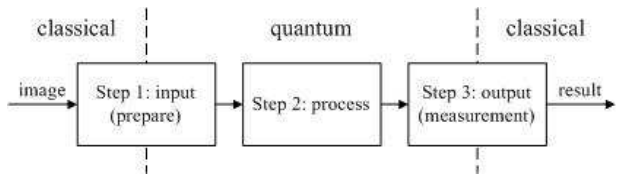}
  \caption{The processing steps of quantum image processing.}
  \label{fig:1}
\end{figure}

Quantum preparation is related to how an image is stored in a
quantum system, which is known as quantum representation. Different
quantum image representations may have different preparation
methods. Researchers have proposed a number of quantum image
representation schemes, such as Qubit Lattice [1], Real Ket [2],
Entangled Image [3], FRQI (the flexible representation of quantum
images) [4], MCQI (the RGB multi-channel representation for quantum
images) [5], NEQR (the novel enhanced quantum representation of
digital images) [6], QUALPI (the quantum representation for
log-polar images) [7], QSMC$\&$QSNC (quantum states for $M$ colors
and quantum states for $N$ coordinates) [8], NAQSS (the normal
arbitrary quantum superposition state) [9], INEQR (the improved
NEQR) [10], GQIR (the generalized quantum image representation)[11],
and \emph{etc}. A number of papers have summarized and compared them
[11-12].

In this paper, we discuss GQIR's compression. GQIR is developed from
FRQI. FRQI, NEQR, INEQR and GQIR belong to one class. They are
approbated and used more commonly. Their feature is that they use
two entangled state sequences to represent color information and
location information respectively. Fig. 2 gives the development and
the inheritance of this class of representation.
\begin{figure}[h]
  \centering
  \includegraphics[width=11cm,keepaspectratio]{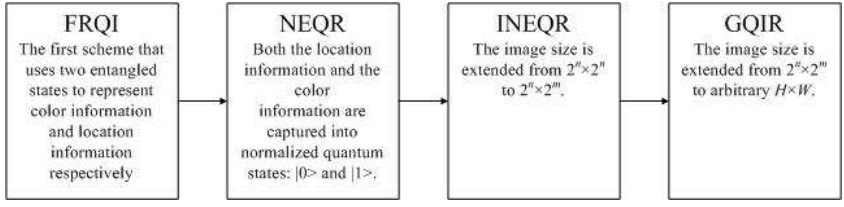}
  \caption{The development and the inheritance of the main class of quantum image representation.}
  \label{fig:2}
\end{figure}

GQIR uses $h=\lceil \log_2 H \rceil$ qubits for $Y$-coordinate and
$w=\lceil \log_2 W \rceil$ qubits for $X$-coordinate to represent a
$H\times W$ image. Both the location information and the color
information are captured into normalized quantum states: $|0\rangle$
and $|1\rangle$. Hence, an GQIR image can be written as

\begin{equation}
    \begin{aligned}
    & |I\rangle = \frac{1}{\sqrt2^{h+w}} \left(\sum_{Y=0}^{H-1}   \sum_{X=0}^{W-1}   |C_{YX} \rangle |YX \rangle \right) \\
    & |YX\rangle=|Y\rangle |X\rangle= |y_{h-1} y_{h-2} \ldots y_{0}\rangle
    |x_{w-1} x_{w-2} \ldots x_{0}\rangle, \ y_{i},x_{i} \in
    \{0,1\}\\
    & |C_{YX}\rangle=|C^{q-1}_{YX}C^{q-2}_{YX}\ldots
    C^{0}_{YX}\rangle, C^{i}_{YX}\in\{0,1\}
    \end{aligned}
\end{equation}
where, $q$ is the color depth, and
\begin{equation*}
 h=\left\{
 \begin{array}{ll}
 \lceil \log_2 H \rceil, & H > 1 \\
 1, & H=1
 \end{array}
 \right.
\end{equation*}
\begin{equation*}
 w=\left\{
 \begin{array}{ll}
 \lceil \log_2 W \rceil, & W > 1 \\
 1, & W=1
 \end{array}
 \right.
\end{equation*}
$|YX\rangle$ is the location information and $|C_{YX}\rangle$ is the
color information. It needs $h+w+q$ qubits to represent a $H\times
W$ image with gray range $2^{q}$. Note that GQIR can represent not
only gray scale images but also color images because the color depth
$q$ is a variable. In most cases, when $q=2$, it is a binary image;
when $q=8$, it is a gray scale image; and when $q=24$, it is a color
image.

We choose GQIR procedure because:
\begin{enumerate}[$\bullet$]
  \item It fully exploits the physical properties of
qubits (entanglement and superposition) and reduces the number of
qubits used to store an image.
  \item It resolves the real-time computation problem of image
processing and provides a flexible method to process any part of a
quantum image by using controlled quantum logic gates.
  \item It can represent a quantum image with any size.
  \item It is close to the classical image representation. Hence, it
  is easier for researchers to understand and to transplant classical image processing
  algorithm into quantum system.
\end{enumerate}

There are also some researchers believe that measurement is a fatal
shortcoming of GQIR because measuring the quantum image one time,
only one pixel is retrieved and the qubits collapse to the pixel,
i.e., other pixels are disappeared. If a user want to retrieve the
whole image, one must prepare, process and measure the image many,
many times. However, we believe this is due to the wrong use, rather
than GQIR itself. It is suitable to solve the problems with small
amount output and can not be solved efficiently in classical
computers. In Ref. [13-14], we have given a more detailed exposition
and two example algorithms about this issue.

Quantum image compression is the procedure that reduces the quantum
resources used to prepare quantum images [4]. The main resource in
quantum preparation is the number of quantum gates instead of the
number of qubits, because as stated previously, the number of qubits
used in quantum images (i.e., $h+w+q$) has been very fewer than the
number of bits used in classical images (i.e., $HWq$). Furthermore,
the number of quantum gates can be used to indicate the network time
complexity because in quantum network, each quantum gate is an
operation which needs a certain amount of time to do it. Hence, in
this paper, we only study network time complexity and simply refer
to as network complexity, or complexity in the following. Therefore,
the main task of quantum image compression is to reduce the number
of gates used during quantum image preparation.

In Ref. [4] and [6], the minimization of Boolean expressions is used
to compress the image preparation. We call it Boolean expression
compression (BEC) method and will introduce it in detail in Section
2.1. However, BEC has some defects that prompt us to give a new
compression scheme: 1) it needs a time-consuming preprocessing; and
2) the compression ratio is unstable.

Hence, this paper gives a new compression scheme. The new one is
based on JPEG which is the most widely used method for still image
compression in classical computers. We call the new one as quantum
image JPEG compression, or the JPEG scheme. It inputs the quantized
JPEG coefficients into qubits and then convert them into pixel
values. Since the data amount of the JPEG coefficients are
significantly less than that of the pixel values, the JPEG scheme
can compress an image. Theoretical analysis and experimental results
show that the compression ratio of our scheme is obviously higher
than that of BEC.

The rest of the paper is organized as follows. Sect. 2 presents
related works about BEC, JPEG and two quantum modules used in our
scheme. Our scheme is discussed in Sect. 3 including the basic ideas
and the scheme steps. Sect. 4 gives the theoretical analysis and
experimental results. Sect. 5 gives the conclusion.

\section{Related works}

\subsection{GQIR preparation and compression}

A GQIR's preparation is composed of $h+w$ Hadamard gates and some
($h+w$)-CNOT gates (CNOT gates with ($h+w$) control qubits). $h+w$
Hadamard gates are used to let state $|0\rangle$ and state
$|1\rangle$ appear with equal probability in $|YX\rangle$, and
($h+w$)-CNOT gates are used to set the color information and
entangle the color information and location information.

In Fig. 3 (a)-(c), an image with 4 pixels is given as an example to
explain GQIR representation and preparation. In (a), one square
indicates one pixel and the number in each square is the decimal
pixel value. In (b), the number in $|\ \rangle$ before each
$\otimes$ is the binary pixel value and the number in $|\ \rangle$
after each $\otimes$ is the binary location information. There are 4
such formulas superposed because the image has 4 pixels. In (c), 2
Hadamard gates generate the 4 location information
$|00\rangle,|01\rangle,|10\rangle,|11\rangle$ and 10 2-CNOT gates
are used to set the color information. Since all of the color
qubits' initial state is $|0\rangle$, the number of CNOT gates is
equal to the number of ``1'' in the color information.

BEC method has been used to compress GQIR, which is based on the
minimization of Boolean expressions. If $x$ is a Boolean variable
and the value of it is 1 then the lateral $x$ is used in the
minterm, otherwise the lateral $\overline{x}$ is used. Then, for
example, a Boolean expression
$$
e=x_{2}\overline{x_{1}}x_{0}+x_{2}x_{1}x_{0}
$$
is minimized as
$$
e=x_{2}x_{0}.
$$
That is to say, if two CNOT gates operate the same color qubit and
only one of their control qubits is different, the two CNOT gates
can be combined into one gate. For example, in Fig. 3(b), the
subspace
\begin{equation*}
\begin{split}
|C^{7}y_{0}x_{0}\rangle&=\frac{1}{2}(|0\rangle\otimes|00\rangle+|1\rangle\otimes|01\rangle+|1\rangle\otimes|10\rangle+|1\rangle\otimes|11\rangle)\\
&=\frac{1}{2}(|0\rangle\otimes|00\rangle+|1\rangle\otimes|01\rangle+|1\rangle\otimes|1\rangle(|0\rangle+|1\rangle))
\end{split}
\end{equation*}
In the first line, $C^{7}$ has three $|1\rangle$, i.e., three CNOT
gates are needed; in the second line, because the last two items are
combined, $C^{7}$ has two $|1\rangle$ and only two CNOT gates are
needed. Fig. 3(d) compresses the whole example image. It can be seen
that BEC drops the number of CNOT gates from 10 to 6.
\begin{figure}[h]
  \centering
  \includegraphics[width=11cm,keepaspectratio]{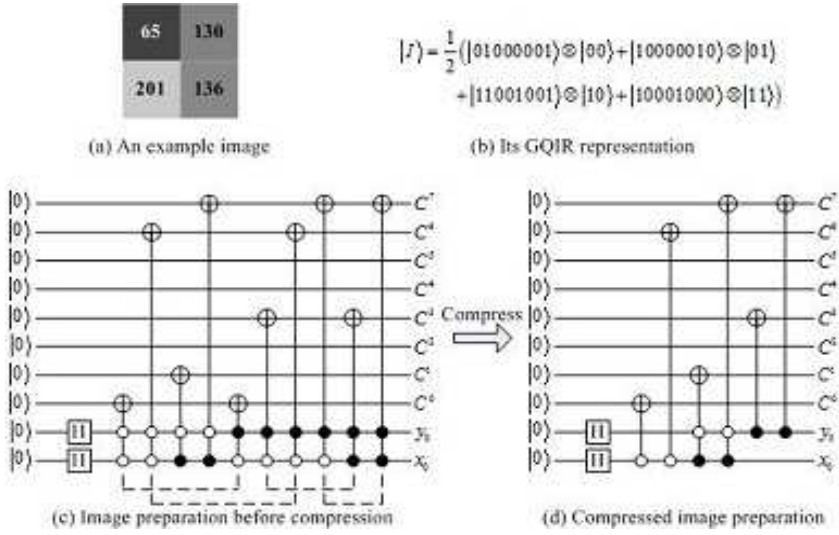}
  \caption{GQIR representation, preparation and compression.}
  \label{fig:3}
\end{figure}

However, the BEC method has some defects:
\begin{enumerate}
\item In fact, BEC has two steps: 1) preprocess: determining which CNOT gates can be combined;
and 2) input: using Hadamard gates and CNOT gates to input the image into quantum system.
The time complexity of BEC is
\begin{equation}
C_{\text{BEC}}=C_{\text{p}}+C_{\text{i}}
\end{equation}
where $C_{\text{p}}$ is the complexity of preprocessing and
$C_{\text{i}}$ is the complexity of inputting. The preprocessing is
time consuming because we have to look over all the CNOT gates
$C_{YX}^{i}$ before compression one pair by one pair. The looking
over algorithm is shown below.
\begin{algorithm}[htb]
  \caption{The looking over algorithm}
  \begin{algorithmic}[1]
    \FORALL {$i=0$ to $(q-1)$}
      \FORALL {$Y_{1}X_{1}=0$ to $(2^{2n}-1)$}
        \FORALL {$Y_{2}X_{2}=Y_{1}X_{1}+1$ to $(2^{2n}-1)$}
          \IF {$C_{Y_{1}X_{1}}^{i}=C_{Y_{2}X_{2}}^{i}=1$ and only one bit of the binary $Y_{1}X_{1}$ and the binary $Y_{2}X_{2}$ is different}
            \STATE Combine the two CNOT gate $C_{Y_{1}X_{1}}^{i}=1$ and $C_{Y_{2}X_{2}}^{i}=1$ to one;
          \ENDIF;
        \ENDFOR;
      \ENDFOR;
    \ENDFOR;
  \end{algorithmic}
\end{algorithm}

Hence, the complexity of the looking over algorithm is
$O(q\cdot2^{4n})$.

Moreover, looking over one round is not enough because two combined
gates may be combined further. Combining two CNOT gate one time, the
number of control qubits will reduce one. Hence, the looking over
process should be executed $2n$ round. Therefore, to a
$2^{n}\times2^{n}$ image, the complexity of preprocessing is
\begin{equation}
C_{\text{p}}=O(2n\cdot q\cdot2^{4n}).
\end{equation}
It is a very high complexity, and in most application scenarios, it
is intolerable. Moreover, it is even much higher than the complexity
before compression: $O(q2^{2n})$. Hence, even assuming that
$C_{\text{i}}$ is 0, BEC method increases the complexity instead of
playing the role of compression.
\item The compression ratio of this method is influenced by the bit
planes. It is difficult to give an equation to calculate the
compression ratio. Sometimes, the compression ratio is high (for an
extreme example, all the pixels have the same color); and sometimes,
maybe, no one CNOT gate can be compressed. That is to say,
$C_{\text{i}}$ has no exact value.
\end{enumerate}

For more information about GQIR, please refer Ref. [11].

\subsection{JPEG compression [15-16]}

JPEG is a commonly used method of lossy compression for digital
still images. The degree of compression can be adjusted, allowing a
selectable tradeoff between storage size and image quality. The
typically compression ratio is 10:1 with little perceptible loss in
image quality. The term ``JPEG'' is an acronym for the Joint
Photographic Experts Group, which created the standard.

\subsubsection{JPEG compression}

The JPEG encoding process consists of the following steps (taking a
gray scale image as an example):
\begin{enumerate}

\item Discrete cosine transform

The image is split into blocks of $8\times8$ pixels, and each block
(denoted as $f(i,j)$, $i=0,1,\cdots,7$, $j=0,1,\cdots,7$) undergoes
the discrete cosine transform (DCT) to get frequency spectrum
$F(u,v)$.
\begin{equation}
F(u,v)=c(u)c(v)\sum_{i=0}^{7}\sum_{j=0}^{7}\left\{f(i,j)\cos\left[\frac{(i+0.5)\pi}{8}u\right]\cos\left[\frac{(j+0.5)\pi}{8}v\right]\right\}
\end{equation}
where $u=0,1,\cdots,7$, $v=0,1,\cdots,7$, and
$$
c(u)=\left\{
\begin{array}{c}
\frac{1}{2\sqrt{2}},u=0\\
\frac{1}{2},u\neq0\\
\end{array}
\right.
$$
$F(0,0)$ is the direct-current coefficient, and the bigger $u$ and
$v$, the higher frequency components $F(u,v)$.

\item Quantization

The human eye is good at seeing small differences in brightness over
a relatively large area, but not so good at distinguishing the exact
strength of a high frequency brightness variation. This allows one
to greatly reduce the amount of information in the high frequency
components. This is done by simply dividing each component in the
frequency domain by a constant for that component, and then rounding
to the nearest integer. This rounding operation is the only lossy
operation in the whole process if the DCT computation is performed
with sufficiently high precision. As a result of this, it is
typically the case that many of the higher frequency components are
rounded to zero, and many of the rest become small positive or
negative numbers, which take many fewer bits to represent. The
elements in the quantization matrix control the compression ratio,
with larger values producing greater compression. A typical
quantization matrix is as follows:
\begin{equation}
Q=\left[
\begin{array}{ccccccccccccccc}
    16 &\ & 11 &\ & 10 &\ & 16 &\ & 24 &\ & 40 &\ & 51 &\ & 61\\
    12 &\ & 12 &\ & 14 &\ & 19 &\ & 26 &\ & 58 &\ & 60 &\ & 55\\
    14 &\ & 13 &\ & 16 &\ & 24 &\ & 40 &\ & 57 &\ & 69 &\ & 56\\
    14 &\ & 17 &\ & 22 &\ & 29 &\ & 51 &\ & 87 &\ & 80 &\ & 62\\
    18 &\ & 22 &\ & 37 &\ & 56 &\ & 68 &\ & 109 &\ & 103 &\ & 77\\
    24 &\ & 35 &\ & 55 &\ & 64 &\ & 81 &\ & 104 &\ & 113 &\ & 92\\
    49 &\ & 64 &\ & 78 &\ & 87 &\ & 103 &\ & 121 &\ & 120 &\ & 101\\
    72 &\ & 92 &\ & 95 &\ & 98 &\ & 112 &\ & 100 &\ & 103 &\ & 99\\
\end{array}
\right]
\end{equation}

The quantized DCT coefficients are computed with
\begin{equation}
F_{Q}(u,v)=round\left(\frac{F(u,v)}{Q(u,v)}\right)
\end{equation}

\item Zigzag

Because so many coefficients in the DCT image are truncated to zero
values during the coefficient quantization stage, the zeros are
handled differently than non-zero coefficients. They are coded using
a Run-Length Encoding (RLE) algorithm. RLE gives a count of
consecutive zero values in the image, and the longer the runs of
zeros, the greater the compression. One way to increase the length
of runs is to reorder the coefficients in the zigzag sequence shown
in Fig. 4. This way, coefficients 0 are omitted and a terminator
``EOF'' is used.
\begin{figure}[h]
  \centering
  \includegraphics[height=5cm,keepaspectratio]{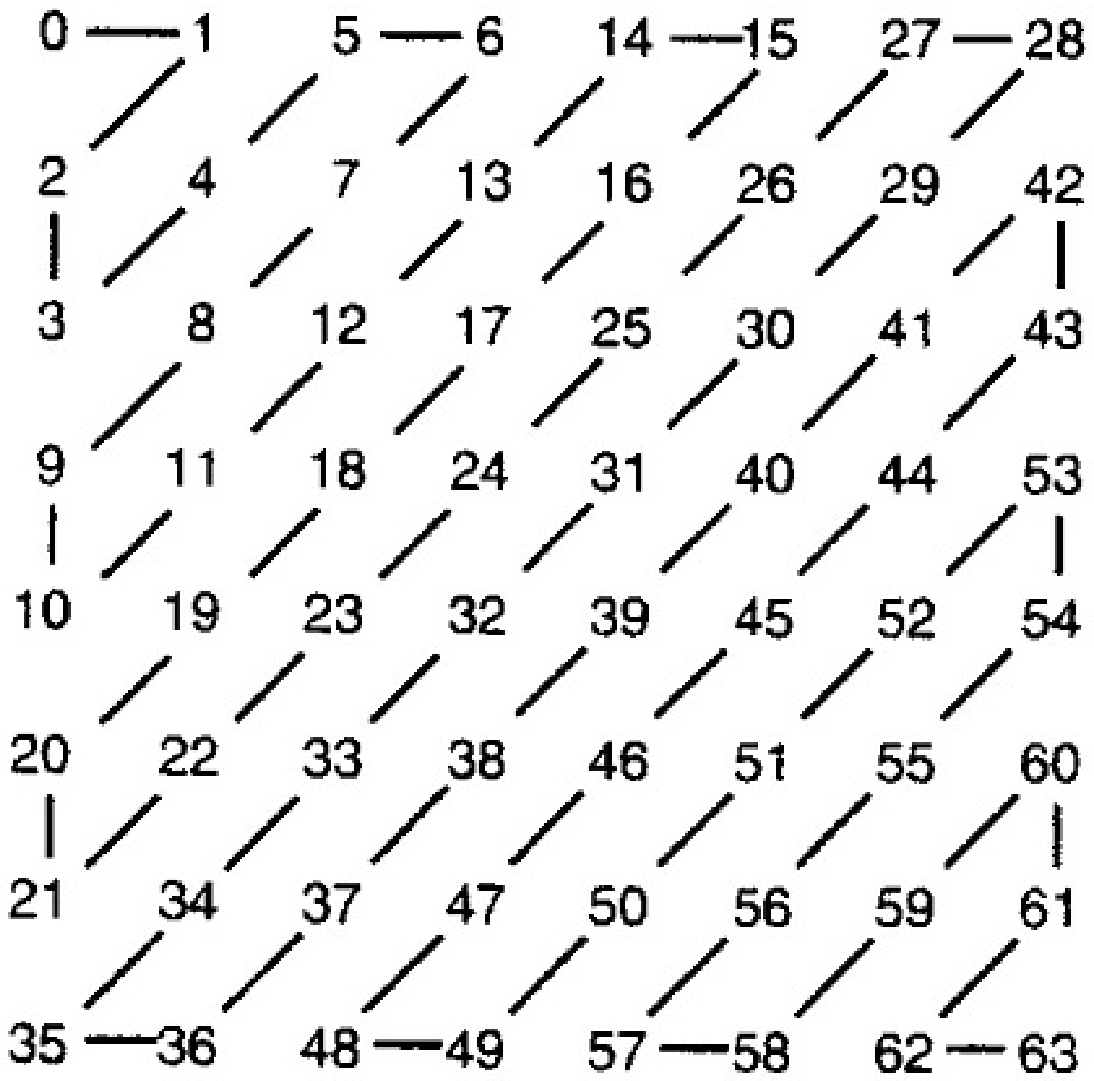}
  \caption{Zigzag.}
  \label{fig:4}
\end{figure}
\end{enumerate}

\subsubsection{JPEG decompression}

If we want to resume the pixel values, for example if we want to
display the image on the screen, the inverse process is executed:
\begin{enumerate}
\item Rearrange the compressed data to $8\times8$ blocks.
\item Multiply each $8\times8$ block with the quantization matrix $Q$.
\begin{equation}
F'(u,v)=F_{Q}(u,v)\times Q(u,v)
\end{equation}
\item Do inverse DCT (IDCT) to each $8\times8$ block to get the
recovered pixel value $f'(i,j)$:
\begin{equation}
f'(i,j)=\sum_{u=0}^{7}\sum_{v=0}^{7}\left\{c(u)c(v)F'(u,v)\cos\left[\frac{(i+0.5)\pi}{8}u\right]\cos\left[\frac{(j+0.5)\pi}{8}v\right]\right\}
\end{equation}
where $i=0,1,\cdots,7$, $j=0,1,\cdots,7$, and
$$
c(u)=\left\{
\begin{array}{c}
\frac{1}{2\sqrt{2}},u=0\\
\frac{1}{2},u\neq0\\
\end{array}
\right.
$$
\end{enumerate}

Fig. 5 shows the JPEG compression of two blocks and the JPEG
decompression of one block in image ``Lena''. Fig. 5 tells us that
the recovered block is not the same as the original block. Hence,
JPEG compression is a kind of lossy compression.
\begin{figure}[h]
  \centering
  \includegraphics[width=12cm,keepaspectratio]{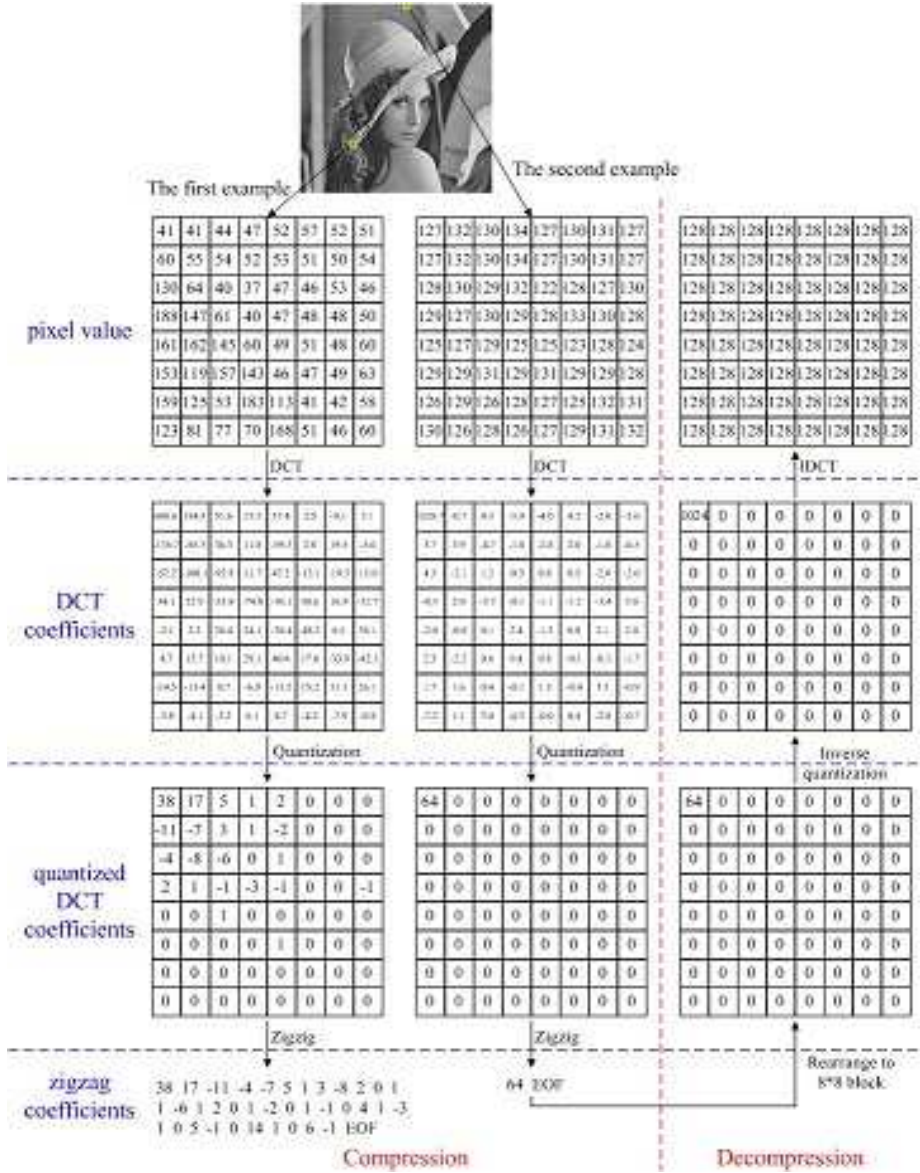}
  \caption{Two examples about the JPEG compression.}
  \label{fig:5}
\end{figure}

\subsection{Quantum multiplier and quantum adder}

Section 2.2 tells us that two operations: multiplication and
addition are used in JPEG compression and decompression. Hence, we
use the quantum multiplier [17] and the quantum adder [18] to
realize them respectively.

The quantum multiplier and the quantum adder are both quantum
networks, which can calculate the product or the sum of two numbers
which are stored in two $n$-qubit quantum registers $|a\rangle$ and
$|b\rangle$, where $a=|a_{n-1}a_{n-2}\cdots a_{0}\rangle$ and
$b=|b_{n-1}b_{n-2}\cdots b_{0}\rangle$.

The quantum multiplication operation $M$ is
\begin{equation}
M(|a\rangle,|b\rangle,|0\rangle^{\otimes
2n})=(|a\rangle,|b\rangle,|a\times b\rangle)
\end{equation}
and the quantum addition operation $A$ is
\begin{equation}
A(|a\rangle,|b\rangle)=(|a\rangle,|a+b\rangle)
\end{equation}
where, $|a\times b\rangle$ is a $2n$-qubit quantum register and
$|a+b\rangle$ is a $(n+1)$-qubit quantum register.

In the following, we use ``MULER'' and ``ADDER'' to represent the
quantum multiplier and the quantum adder respectively as shown in
Fig. 6.
\begin{figure}[h]
  \centering
  \subfigure[The quantum multiplier]{
    \label{fig:subfig:a}
    \includegraphics[height=3cm]{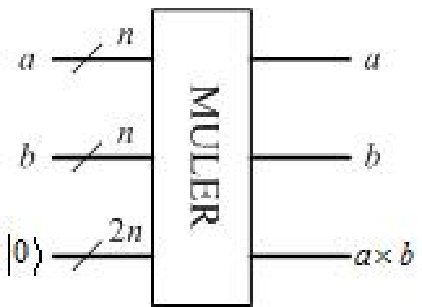}}
  \hspace{0.2in}
  \subfigure[The quantum adder]{
    \label{fig:subfig:b}
    \includegraphics[height=3cm]{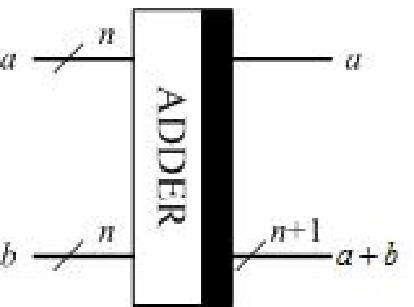}}
  \caption{The quantum multiplier and the quantum adder.}
  \label{fig:6}
\end{figure}

According to Ref. [17] and [18], the complexity of MULER is
\begin{equation}
\begin{split}
 &n^{2}\text{ (2)-CNOT gates}+\sum_{l=1}^{\log_{2}n}\text{the complexity of Stage }l\\
=&n^{2}+\sum_{l=1}^{\log_{2}n}(4n+4l+4\cdot2^{l-1}-10)\\
=&n^{2}+4n-4+(4n-8+2\log_{2}n)\log_{2}n
\end{split}
\end{equation}
The complexity of ADDER is
\begin{equation}
\begin{split}
 &(2n-1)\text{ Carry modules}+n\text{ Sum modules}+1\text{ CNOT gate}\\
=&3(2n-1)+2n+1\\
=&8n-2
\end{split}
\end{equation}

Note that
\begin{enumerate}[$\bullet$]
\item In Fig. 6(b), there has a thick black bar on the right-hand side of the
quantum adder. A network with a bar on the left side represents the
reversed sequence of elementary gates embedded in the same network
with the bar on the right side. If the action of the quantum adder
is reversed with the input $(a,b)$, the output will output $(a,b-a)$
when $b\geq a$. When $b<a$, the output is $(a,2^{n}-(a-b))$.
\item Ref. [17] said that Stage $l$ had $\frac{n}{2^{l}}$ ripple quantum
adders with size $n+l+2^{l-1}-2$ in parallel. Since the
$\frac{n}{2^{l}}$ ripple quantum adders work at the same time, the
time required to run all the $\frac{n}{2^{l}}$ ripple quantum adders
is equal to the run time of one ripple quantum adder. That is to say
the time complexity of State $l$ is equal to one ripple quantum
adder.
\item The ripple quantum adder is different from the quantum adder given in Ref. [18]. The time complexity of a ripple quantum adder with size $n$ is $4n-2$. Hence, the complexity of a ripple quantum adder with size $n+l+2^{l-1}-2$ is $4n+4l+4\cdot2^{l-1}-10$.
\item Due to the length limitation of this paper, to get more
details about MULER and ADDER, please refer to Ref. [17] and [18].
\end{enumerate}

\section{Quantum JPEG image compression}

\subsection{Basic idea}

Unlike the existed schemes that store pixel values in quantum
computers directly, the basic idea of our scheme is to store
quantized DCT coefficients into qubits and then transform them into
pixel values. Due to the volume of DCT coefficients is obviously
smaller than the volume of pixel values, the complexity of
preparation is decreased, i.e., the quantum image is compressed.
Fig. 7 gives a schematic diagram to contrast our scheme with
traditional schemes.

\begin{figure}[h]
  \centering
  \includegraphics[width=11cm,keepaspectratio]{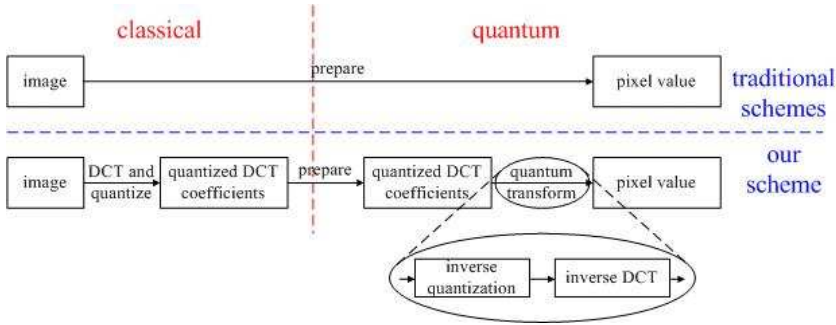}
  \caption{The basic idea of our scheme and the contrast with traditional schemes.}
  \label{fig:7}
\end{figure}

Note that the zigzag step is no longer needed in our scheme because
the quantized coefficients are input into quantum computers in the
form of $8\times8$ blocks directly.

Section 2.2.2 tells us that when transform the quantized DCT
coefficients into pixel values, it needs two steps: multiply each
$8\times8$ block with the quantization matrix $Q$ (inverse
quantization), and do inverse DCT (IDCT) to each $8\times8$ block to
get the recovered pixel value $f'(i,j)$.

Based on the unitarity of quantum algorithms, quantum IDCT can be
realized by simply reverse the order of logic gates in quantum DCT
algorithms. However the existed quantum DCT algorithms [19-20] is
not suitable to our scheme because their data storage format is
different from ours:
\begin{enumerate}[$\bullet$]
\item Ref. [19-20] use $d=Uc$ to do quantum DCT, where $c=[c_{0}\ c_{1}\ \cdots\
c_{2^{n}-1}]^{\text{T}}$ is the space domain data (i.e., the
original pixel values), $d=[d_{0}\ d_{1}\ \cdots\
d_{2^{n}-1}]^{\text{T}}$ is the frequency domain data (i.e., the DCT
coefficients), and $U$ is the quantum DCT transformation. Hence, the
pixels and the coefficients are stored as the probability of basic
states.
\item Our scheme converts the pixel values and the coefficients to binary
strings and use the corresponding basic states to represent them.
The probabilities of all the corresponding basic states are the
same.
\end{enumerate}

Hence, the quantum DCT and IDCT method proposed in Ref. [19-20] can
not be utilized in this paper. We will give a new way to realize
quantum IDCT and describe it in the next subsection.

\subsection{Quantum JPEG image compression algorithm}

For the sake of simplicity, we assume that the size of the image is
$2^{n}\times2^{n}$.

\begin{enumerate}[Step 1]
\item DCT and quantize.
\end{enumerate}

It is a preprocessing step done in classical computers as described
in Step 1 and 2 of Section 2.2.1.

\begin{enumerate}[Step 2]
\item Store the quantized DCT coefficients into quantum systems.
\end{enumerate}

It is similar to the GQIR preparation stated in Section 2.1.

Initially, $q+2n$ qubits are prepared and set to $|0\rangle$. The
initial state can be expressed as in
\begin{equation}
|\Psi_{0}\rangle=|0\rangle^{\otimes q+2n}
\end{equation}

Then, $q$ identity gates and $2n$ Hadamard gates are used to
construct a $2^{n}\times2^{n}$ blank image. The identity matrix and
the Hadamard matrix are shown below.
 \begin{equation}
 I=\left[
 \begin{array}{cc}
 1 & 0 \\
 0 & 1
 \end{array}
 \right]
 \ \ \ \
 H=\frac{1}{\sqrt2} \left[
 \begin{array}{cc}
 1 & 1 \\
 1 & -1
 \end{array}
 \right]
 \end{equation}

The whole quantum operation in this step can be expressed as
$U_{2.1}$:
\begin{equation}
U_{2.1}=I^{\otimes q}\otimes H^{\otimes 2n}
\end{equation}
$U_{2.1}$ transforms the initial state $|\Psi_{0}\rangle$ to the
state $|\Psi_{1}\rangle$.
\begin{equation}
\begin{split}
|\Psi_{1}\rangle&=U_{2.1}(|\Psi_{0}\rangle)=(I|0\rangle)^{\otimes
q}\otimes (H|0\rangle)^{\otimes 2n}\\
&=\frac{1}{2^{n}}|0\rangle^{\otimes
q}\otimes\sum_{i=0}^{2^{2n}-1}|i\rangle\\
&=\frac{1}{2^{n}}\sum_{Y=0}^{2^{n}-1}\sum_{X=0}^{2^{n}-1}|0\rangle^{\otimes
q}|YX\rangle
\end{split}
\end{equation}

In simple terms, the effect of identity gate is maintaining the
qubit's original state unchanged and the effect of Hadamard gate is
letting state $\vert 0 \rangle$ and state $\vert 1 \rangle$ appear
with equal probability. Note that because identity gates $I$ remain
qubit's state unchanged, they can be omitted. After $U_{2.1}$, the
blank $2^{n}\times2^{n}$ image is gained.

Next, store the quantized DCT coefficients. It is divided into
$2^{n-3}\times2^{n-3}\times64=2^{2n}$ sub-operations and each
sub-operation stores one quantized DCT coefficient, where
$2^{n-3}\times2^{n-3}$ is the number of $8\times8$ blocks and 64
indicates that there are 64 coefficients to be stored in each block.
We still use $X$ and $Y$ to distinguish sub-operations. For the
$(u,v)$th sub-operation in block $(i,j)$, we have
\begin{equation}
\begin{split}
&X=8j+v\\
&Y=8i+u
\end{split}
\end{equation}

In general, in each block, $F_{Q}(0,0)$ is the biggest coefficient
of all $F_{Q}(u,v)$, $u=0,\cdots,7$, $v=0,\cdots,7$. Due to
$$
F_{Q}(0,0)=round\left(\frac{F(0,0)}{Q(0,0)}\right)=round\left(\frac{1}{16}\times\frac{1}{8}\sum_{i=0}^{7}\sum_{j=0}^{7}f(i,j)\right)
$$
and $f(i,j)$ does not exceed $2^{q}-1$,
$$
F_{Q}(0,0)<
round\left(\frac{1}{16}\times\frac{1}{8}\times64\left(2^{q}-1\right)\right)=2^{q-1}.
$$
Note that there is a special item. When all $f(i,j)=2^{q}-1$,
$$round\left(\frac{1}{16}\times\frac{1}{8}\times64\left(2^{q}-1\right)\right)=round(2^{q-1}-0.5)=2^{q-1}.$$
However, JPEG compression defines that the quantization result of
the special item is $2^{q-1}-1$.

Hence, $q-1$ qubits are enough to store $F_{Q}(u,v)$. However,
sometimes, $F_{Q}(u,v)$ may be a negative number. The most
significant qubit is used as the sign position: if $F_{Q}(u,v)$ is a
negative number, the most significant qubit is set to state
$|1\rangle$; and if $F_{Q}(u,v)$ is a positive number, the most
significant qubit remains state $|0\rangle$.

The quantum sub-operation $U_{YX}$ is shown
 below.
 \begin{equation}
 U_{YX}=\left(I\otimes\sum_{ji\neq YX}|ji\rangle\langle
 ji|\right)+\Omega_{YX}\otimes|YX\rangle\langle
 YX|
 \end{equation}
where $\Omega_{YX}$ is the quantum operation to transform the value
of pixel $(Y,X)$ from $|0\rangle^{\otimes q}$ to the desired
quantized DCT coefficient as shown in Eq. (19). Hence, the function
of $U_{YX}$ is change pixel $(Y,X)$ and others remain unchanged.
\begin{equation}
\Omega_{YX}=\bigotimes_{i=0}^{q-1}\Omega_{YX}^{i}
\end{equation}
The function of $\Omega_{YX}^{i}$ is setting the value of the $i$th
qubit of pixel $(Y,X)$'s quantized DCT coefficient.
\begin{equation}
\Omega_{YX}^{i}:|0\rangle\rightarrow|0\oplus C_{YX}^{i}\rangle
\end{equation}
where $\oplus$ is the XOR operation. If $C_{YX}^{i}=1$,
$\Omega_{YX}^{i}:|0\rangle\rightarrow|1\rangle$ is a $2n$-CNOT gate
(a CNOT gate with $2n$ control qubits). Otherwise,
$\Omega_{YX}^{i}:|0\rangle\rightarrow|0\rangle$ is a quantum
identity gate which will do nothing on the quantum state. Hence,
\begin{equation}
\Omega_{YX}|0\rangle^{\otimes
q}=\bigotimes_{i=0}^{q-1}(\Omega_{YX}^{i}|0\rangle)=\bigotimes_{i=0}^{q-1}|0\oplus
C_{YX}^{i}\rangle=\bigotimes_{i=0}^{q-1}|C_{YX}^{i}\rangle=|C_{YX}\rangle
\end{equation}
Act $U_{YX}$ on $|\Psi_{1}\rangle$
\begin{equation}
\begin{split}
U_{YX}(|\Psi_{1}\rangle)&=U_{YX}\left(\frac{1}{2^{n}}\sum_{j=0}^{2^{n}-1}\sum_{i=0}^{2^{n}-1}|0\rangle^{\otimes
q}|ji\rangle\right)\\
&=\frac{1}{2^{n}}U_{YX}\left(\sum_{ji\neq YX}|0\rangle^{\otimes
q}|ji\rangle+|0\rangle^{\otimes
q}|YX\rangle\right)\\
&=\frac{1}{2^{n}}\left(\sum_{ji\neq YX}|0\rangle^{\otimes
q}|ji\rangle+\Omega_{YX}|0\rangle^{\otimes
q}|YX\rangle\right)\\
&=\frac{1}{2^{n}}\left(\sum_{ji\neq YX}|0\rangle^{\otimes
q}|ji\rangle+|C_{YX}\rangle|YX\rangle\right)
\end{split}
\end{equation}

$U_{YX}$ only sets the quantized DCT coefficient of its
corresponding pixel. In order to set all the $2^{n}\times2^{n}$
pixels, a quantum operation $U_{2.2}$ is defined below.
\begin{equation}
U_{2.2}=\prod_{\substack{X=0,\cdots,2^{n}-1\\
Y=0,\cdots,2^{n}-1}}U_{YX}
\end{equation}
Act $U_{2.2}$ to $|\Psi_{1}\rangle$
\begin{equation}
|\Psi_{2}\rangle=U_{2.2}(|\Psi_{1}\rangle)=\frac{1}{2^{n}} \sum_{\substack{X=0,\cdots,2^{n}-1\\
Y=0,\cdots,2^{n}-1}} |C_{YX}\rangle |YX \rangle
\end{equation}

In simple terms, the effect of Step 2 is using $2n$ Hadamard gates
and $2n$-CNOT gates to store the quantized DCT coefficients to a
quantum system. The number of $2n$-CNOT gates is equal to the number
of ``1'' in binary coefficients plus the number of negative
coefficients.

\begin{enumerate}[Step 3]
  \item Store the quantization matrix.
\end{enumerate}

This step uses the similar method as the previous step to store the
quantization matrix $Q$ (shown in Eq. (5)) into qubits: 1)
$(q-1)+6=q+5$ qubits are prepared and set to $|0\rangle$. 2) 6
Hadamard gates are used to construct a $8\times8$ blank matrix. 3)
Some 6-CNOT gates are used to set the value of $Q$.

Due to $Q$ is a $8\times8$ matrix, the CNOT gates used in this step
have 6 control qubits. After Step 3, the quantum quantization matrix
is:
\begin{equation}
\begin{split}
|Q\rangle=&\sum_{Y_{Q}=0}^{7}\sum_{X_{Q}=0}^{7}|Q_{Y_{Q}X_{Q}}\rangle\otimes|Y_{Q}X_{Q}\rangle\\
=&|16\rangle|000000\rangle+|11\rangle|000001\rangle+|10\rangle|000010\rangle+|16\rangle|000011\rangle\\
&+\cdots+\\
&+|112\rangle|111100\rangle+|100\rangle|111101\rangle+|103\rangle|111110\rangle+|99\rangle|111111\rangle
\end{split}
\end{equation}
where, in the second to fourth line, $Q_{Y_{Q}X_{Q}}$ is given in
decimal and $Y_{Q}X_{Q}$ is given in binary.

\begin{enumerate}[Step 4]
  \item Inverse quantization.
\end{enumerate}

We use MULER to multiply each $8\times8$ block with the quantization
matrix $|Q\rangle$ as Eq. (7) shows. The multiplication is
element-by-element: only when
$|Y_{Q}X_{Q}\rangle=|Y_{2}Y_{1}Y_{0}X_{2}X_{1}X_{0}\rangle$, MULER
is used to multiply the coefficient block and the quantization
matrix, where $Y_{Q}X_{Q}$ is the location information of the
quantization matrix and $Y_{2}Y_{1}Y_{0}X_{2}X_{1}X_{0}$ is part of
the location information of the DCT coefficient. That is to say,
each element in the quantization matrix is multiplied with the
corresponding DCT coefficients in each $8\times8$ block.

Step 4 can be broken down into the following steps.

\begin{enumerate}[Step 4.1]
  \item Judge whether
  $|Y_{Q}X_{Q}\rangle=|Y_{2}Y_{1}Y_{0}X_{2}X_{1}X_{0}\rangle$.

Define
\begin{equation}
U_{4.1}^{i,T}:|T_{Qi},T_{i}\rangle\rightarrow|T_{Qi}\oplus
T_{i},T_{i}\rangle
\end{equation}
where $T\in\{Y,X\}$, $i\in\{0,1,2\}$. That is to say,
$U_{4.1}^{i,T}$ is a CNOT gate: if $T_{Qi}=T_{i}$, $T_{Qi}$ is
changed to 0; otherwise, $T_{Qi}$ is changed to 1. Hence, by acting
\begin{equation}
U_{4.1}=\bigotimes_{\substack{i=0,1,2\\T=Y,X}}U_{4.1}^{i,T}
\end{equation}
to state $|\Psi_{2}\rangle\otimes|Q\rangle$, we can align $8\times8$
blocks with $|Q\rangle$, i.e., if all the $|T_{Qi}\oplus
T_{i}\rangle$ is equal to $|0\rangle$, each element in the
quantization matrix finds its corresponding DCT coefficients in each
$8\times8$ block. For the sake of simplicity, when no ambiguity is
possible, we still use $|T_{Qi}\rangle$ to substitute $|T_{Qi}\oplus
T_{i}\rangle$.

  \item If the condition is satisfied (i.e., all the $|T_{Qi}\rangle$ is equal to $|0\rangle$), change the state of a auxiliary
qubit $|g\rangle$ with initial state $|0\rangle$ to $|1\rangle$.

A transform $U_{4.2}$ is defined.
\begin{equation}
U_{4.2}:|g=0\rangle\rightarrow|g=1\rangle\text{, if
}|Y_{Q}X_{Q}\rangle=|000000\rangle
\end{equation}
where, $U_{4.2}$ is a 6-CNOT gate.

  \item If $|g=1\rangle$, multiply.

$|g\rangle$ is the deliverable of Step 4.1 and 4.2. Hence,
$|Y_{Q}X_{Q}\rangle$ has no use in the following and $|g\rangle$
substitutes them. If $|g=1\rangle$, we use MULER to multiply each
$8\times8$ block with the quantization matrix $|Q\rangle$ as Eq. (7)
shows. We use $U_{4.3}$ to denote the operation.
\begin{equation}
U_{4.3}:|F,Q,0\rangle\rightarrow|F,Q,F'=F\times Q\rangle\text{, if
}|g\rangle=|1\rangle
\end{equation}

Since the quantization matrix and the DCT coefficients are stored
superposedly, one MULER is enough to multiply them parallel.

  \item Transform the sign position.

Define
\begin{equation}
U_{4.4}:|F'^{2(q-1)-1},F^{q-1}\rangle\rightarrow|F'^{2(q-1)-1}\oplus
F^{q-1},F^{q-1}\rangle\text{, if }|g\rangle=|1\rangle
\end{equation}
Hence, $U_{4.4}$ is a 2-CNOT gate: if $|g\rangle=|1\rangle$ and
$|F^{q-1}\rangle=|1\rangle$, the sign position
$|F'^{2(q-1)-1}\rangle$ in $|F'\rangle$ is changed to $|1\rangle$.
\end{enumerate}

Fig. 8 gives the circuit of Step 4. When $|g\rangle=|1\rangle$, the
subspace $|F'\rangle\otimes|YX\rangle$ stores the recovered DCT
coefficients.
\begin{figure}[h]
  \centering
  \includegraphics[width=11cm,keepaspectratio]{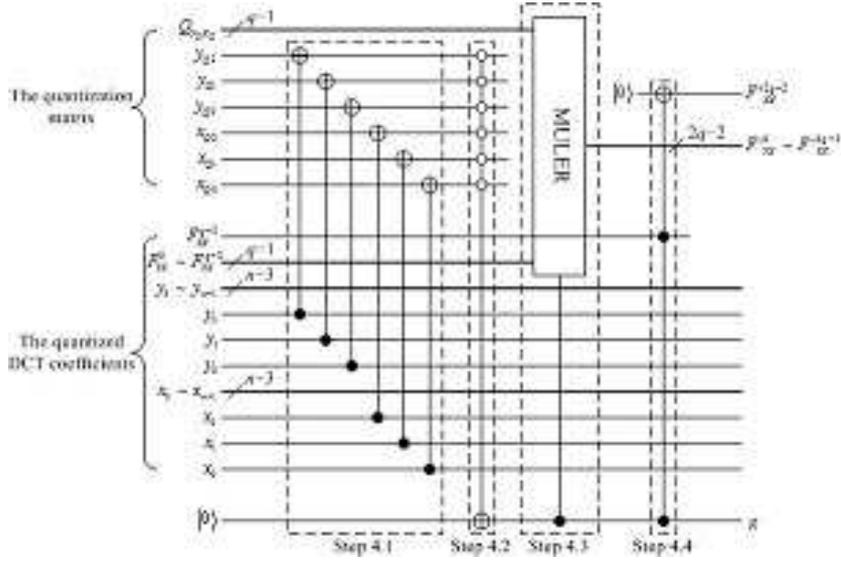}
  \caption{Step 4: inverse quantization.}
  \label{fig:8}
\end{figure}

\begin{enumerate}[Step 5]
  \item Inverse DCT.
\end{enumerate}

In Eq. (8), $c(u)$, $c(v)$, $\cos\left[\frac{(i+0.5)\pi}{8}u\right]$
and $\cos\left[\frac{(j+0.5)\pi}{8}v\right]$ are independent from
image. We see them as a whole. Hence, Eq. (8) can be changed into
\begin{equation}
\begin{split}
f'(i,j)=&\sum_{u=0}^{7}\sum_{v=0}^{7}\left\{c(u)c(v)F'(u,v)\cos\left[\frac{(i+0.5)\pi}{8}u\right]\cos\left[\frac{(j+0.5)\pi}{8}v\right]\right\}\\
=&\sum_{u=0}^{7}\sum_{v=0}^{7}\left\{F'(u,v)\times
C_{ij}(u,v)\right\}
\end{split}
\end{equation}
where
$C_{ij}(u,v)=c(u)c(v)\cos\left[\frac{(i+0.5)\pi}{8}u\right]\cos\left[\frac{(j+0.5)\pi}{8}v\right]$.

Therefore, this step is divided into two main parts: multiply
$F'(u,v)$ with $C_{ij}(u,v)$ and add all the 64 products up.

\begin{enumerate}[Step 5.1]

  \item Store $C_{ij}(u,v)$.

This step uses the similar method as Step 2 to store $C$ into
qubits: 1) $64(q+4)+6=64q+262$ qubits are prepared and set to
$|0\rangle$. 2) 6 Hadamard gates are used to construct $i$ and $j$.
3) Some 6-CNOT gates are used to set the value of $C$.

Although $|F'_{YX}\rangle\otimes|YX\rangle$ use $2q-1$ qubits to
represent the value of DCT coefficients, only $q+4$ qubits works.
That is because in a $8\times8$ block, the max value of $F'(u,v)$ is
$F'(0,0)$, and
$$
F'(0,0)=F_Q(0,0)\times Q(0,0)=16F_Q(0,0)<16\cdot2^{q-1}=2^{q+3}
$$
Therefore we only take the first $q+3$ qubits ($F_{YX}^{'q-5}\sim
F_{YX}^{'2q-3}$) to participate in the following steps.
$F_{YX}^{'2q-2}$ still is the sign. In order to multiply $F'$ with
$C$ using MULER, we also use $q+3$ qubits to store the value of
$C_{ij}(u,v)$ and $C_{ij}^{q+3}(u,v)$ is the sign position.

$C_{ij}(u,v)$ has up to $8^{4}=4096$ values because
$i,j,u,v\in\{0,1,\cdots,6,7\}$. By calculating all of them, we find
that
\begin{equation}
-0.2405\leq C_{ij}(u,v)\leq0.2405
\end{equation}
Hence, except the sign qubit $C_{ij}^{q+3}(u,v)$, other $q+3$ qubits
store the fractional part of the absolute value.

That is to say, each $C_{ij}(u,v)$ uses $q+4$ qubits to store.
However, because $u,v\in\{0,1,2,3,4,5,6,7\}$, $64(q+4)$ qubits are
needed to store all $C_{ij}(u,v)$.

  \item Multiply the absolute value of
$F'(u,v)$ with $C_{ij}(u,v)$.

Define
\begin{equation}
\begin{split}
U_{5.2}^{u,v}:&|F'(u,v),C_{ij}(u,v)\rangle\\
\rightarrow&|P^{2q+5}_{ij}(u,v)\cdots
P^{0}_{ij}(u,v)\rangle=|F'(u,v)\times C_{ij}(u,v)\rangle
\end{split}
\end{equation}
It is a quantum multiplier MULER with 6 control qubits:
$|y_{2}y_{1}y_{0}\rangle$ and $|x_{2}x_{1}x_{0}\rangle$ are the
control qubits, and if the binary values $y_{2}y_{1}y_{0}=u$ and
$x_{2}x_{1}x_{0}=v$, MULER works. Due to the absolute values of
$F'(u,v)$ and $C_{ij}(u,v)$ both occupy $q+3$ qubits, the result
occupies $2q+6$ qubits.

Use 64 MULERs
\begin{equation}
U_{5.2}=\bigotimes_{\substack{u=0,\cdots,7\\v=0,\cdots,7}}U_{5.2}^{u,v}
\end{equation}
to multiply all $F'(u,v)$ with $C_{ij}(u,v)$.

Symbol $P_{ij}(u,v)$ is used to denote the product of $F'(u,v)$ and
$C_{ij}(u,v)$. Since $F'(u,v)$ is a $(q+3)$-qubit integer and
$C_{ij}(u,v)$ is a $(q+3)$-qubit pure decimal, the higher $q+3$
qubits of $P_{ij}(u,v)$ are the integer part and the lower $q+3$
qubits are the fractional part.
  \item Set the sign position.

Define
\begin{equation}
\begin{split}
U_{5.3}^{u,v}:&|0\rangle\rightarrow|1\rangle\text{, if
}y_{2}y_{1}y_{0}=u\text{ and
}x_{2}x_{1}x_{0}=v\\
&\text{ and }((F^{'2q-2}(u,v)=1\text{ and
}C^{q+3}_{ij}(u,v)=0)\\
&\text{ or }(F^{'2q-2}(u,v)=0\text{ and }C^{q+3}_{ij}(u,v)=1))
\end{split}
\end{equation}
It is a 8-CNOT gate: if the sign position of $F'(u,v)$ is different
with the sign position of $C_{ij}(u,v)=1$, and $y_{2}y_{1}y_{0}=u$
and $x_{2}x_{1}x_{0}=v$, change the new sign position
$|P^{2q+6}_{ij}(u,v)\rangle$ from $|0\rangle$ to $|1\rangle$.

Define
\begin{equation}
U_{5.3}=\bigotimes_{\substack{u=0,\cdots,7\\v=0,\cdots,7}}U_{5.3}^{u,v}
\end{equation}
to set all the 64 sign positions.

  \item Add the 64 $|P_{ij}(u,v)\rangle$.

According to Eq. (31), all the 64 $|P_{ij}(u,v)\rangle$ should be
added together. We use 64 ADDER to realize it. Firstly,
$|P^{2q+5}_{ij}(0,0)\cdots P^{0}_{ij}(0,0)\rangle$ and
$|0\rangle^{\otimes 2q+6}$ are added. Then,
$|P^{2q+5}_{ij}(u,v)\cdots P^{0}_{ij}(u,v)\rangle$ and the sum of
the previous addition are added.

However, there is a sign position $|P^{2q+6}_{ij}(u,v)\rangle$. If
it is $|0\rangle$, $|P^{2q+5}_{ij}(u,v)$ $\cdots
P^{0}_{ij}(u,v)\rangle$ should be added; if it is $|1\rangle$,
$|P^{2q+5}_{ij}(u,v)\cdots P^{0}_{ij}(u,v)\rangle$ should be
subtracted. Hence, $|P^{2q+6}_{ij}(u,v)\rangle$ is used as a control
qubit: when $|P^{2q+6}_{ij}(u,v)\rangle=|0\rangle$, the thick black
bar is on the right-hand hand of ADDER; otherwise, the thick black
bar is on the left-hand hand of ADDER.

The output of the last ADDER is the regained color information. It
is denoted as $|f'^{2q+5}\cdots f'^{0}\rangle$ and the higher $q+3$
qubits are the integer part and the lower $q+3$ qubits are the
fractional part. However, GQIR only uses $q$ qubits to represent
color. Hence, only the lower $q$ qubits of the integer part is
useful. It is $|f'^{2q+2}\cdots f'^{q+3}\rangle$.

\end{enumerate}

Fig. 9 gives the circuit of Step 5. The useful output of Step 5 is
$|f'^{2q+2}\cdots$ $f'^{q+3}\rangle\otimes|y_{n-1}\cdots
y_{3}i_{2}i_{1}i_{0}x_{n-1}\cdots x_{3}j_{2}j_{1}j_{0}\rangle$ and
it is the GQIR image. $|f'^{2q+2}\cdots$ $f'^{q+3}\rangle$ is the
color information and $|y_{n-1}\cdots
y_{3}i_{2}i_{1}i_{0}x_{n-1}\cdots x_{3}j_{2}j_{1}j_{0}\rangle$ is
the location information.
\begin{figure}[h]
  \centering
  \includegraphics[width=12cm,keepaspectratio]{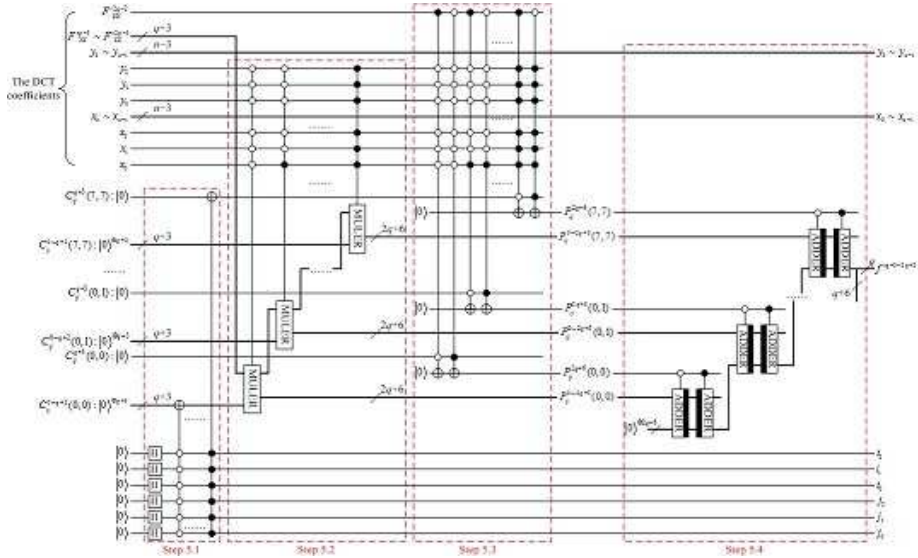}
  \caption{Step 5: inverse DCT.}
  \label{fig:9}
\end{figure}

\section{Network complexity and compression ratio}

In this section, we will compare our scheme with GQIR preparation
and BEC method to highlight the advantages of our scheme. However,
our scheme and BEC have a preprocessing step done in classical
computers. As we all know, classical computers and quantum computers
have different physical principles. Hence, we will compare the
quantum part and the classical part separately.

\subsection{Quantum part}

The network complexity depends very much on what is considered to be
an elementary gate. This is a problem about gate granularity. For
example, one Toffoli gate (a NOT gate with 2 control qubits) can be
simulated by six Control-NOT gates (a NOT gate with 1 control
qubits) [18]. However, in this paper, we take one gate can be drawn
in a quantum circuit as an elementary gate, no matter how many
control qubits it has and what function it finishes.

The GQIR preparation is done in quantum computers. From Section 2.1,
we know that the complexity of the GQIR preparation is equal to the
number of bit 1 in all the pixels. By assuming that every bit has
equal probability to be 0 or 1, then the GQIR preparation complexity
before compression is
\begin{equation}
C=\frac{1}{2}2^{n}\times2^{n}\times q=\frac{q}{2}2^{2n}
\end{equation}
We will give our scheme's complexity based on $C$.

To our scheme, only Step 2-5 are analyzed because the scope of
Section 4.1 is quantum part.

The complexity of Step 2 (The number of gates used in Step 2) in our
scheme is equal to the number of bit 1 plus the number of negatives
in all the coefficients. We define it as
\begin{equation}
r_{J}\cdot C
\end{equation}
where,
\begin{equation}
\begin{split}
r_{J}&=\frac{\text{(The number of bit 1}+\text{the number of
negatives) in all the coefficients}}{\text{The number of bit 1 in
all the pixels}}\\
&=\frac{\text{The complexity after compression}}{\text{The
complexity before compression}}
\end{split}
\end{equation}
is called JPEG compression ratio. We use statistical method to gain
$r_{J}$. 800 images obtained from Washington University [21],
University of South California [22] and Pixabay [23] are used,
including people, animals, streetscape, buildings, natural beauty,
remote sensing, texture, man-made pattern, and so on. These images
almost cover all common image types. Fig. 10 gives some samples.
\begin{figure}[h]
  \centering
  \subfigure[people]{
    \label{fig:subfig:a}
    \includegraphics[height=2cm]{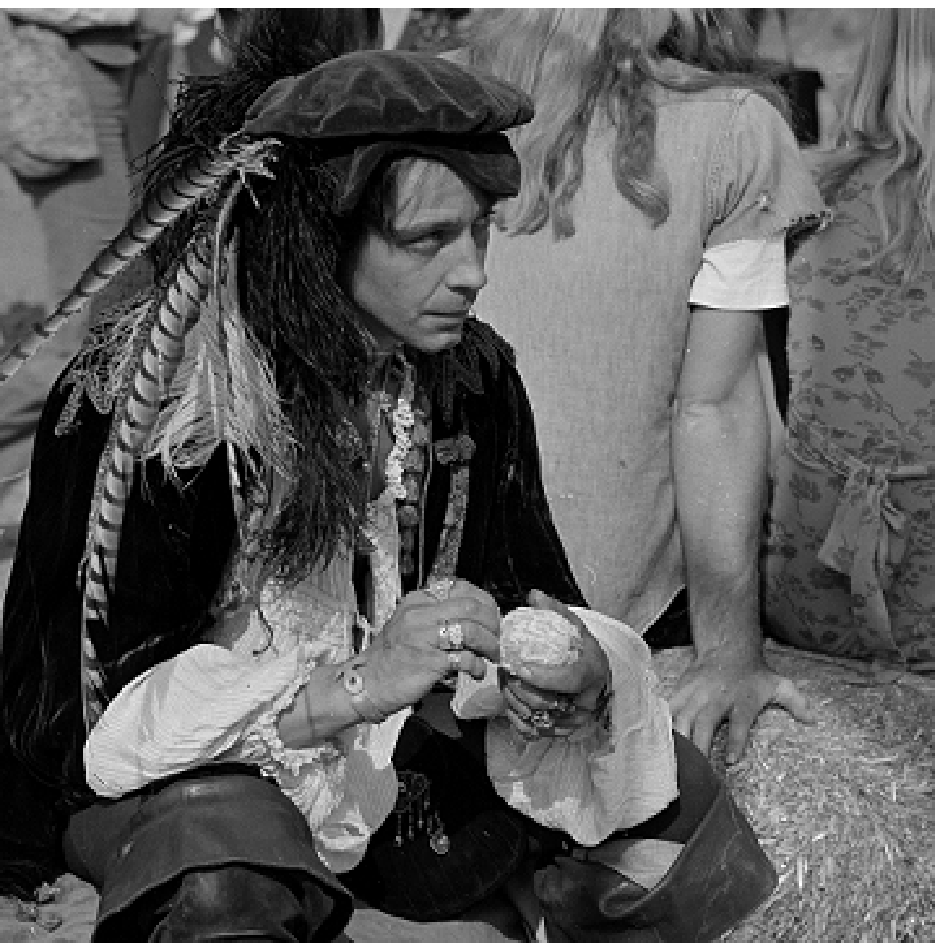}}
  \hspace{0.05in}
  \subfigure[animal]{
    \label{fig:subfig:b}
    \includegraphics[height=2cm]{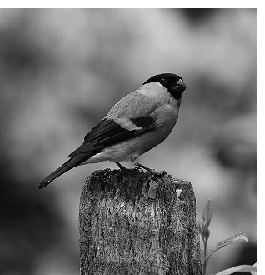}}
  \hspace{0.05in}
  \subfigure[streetscape]{
    \label{fig:subfig:c}
    \includegraphics[height=2cm]{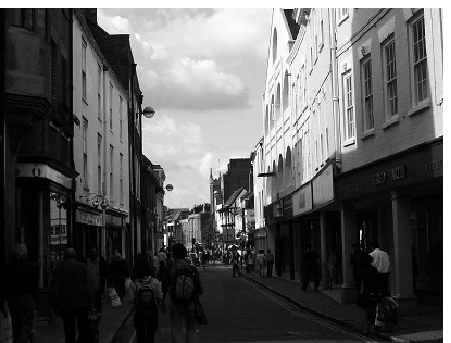}}
  \hspace{0.05in}
  \subfigure[building]{
    \label{fig:subfig:d}
    \includegraphics[height=2cm]{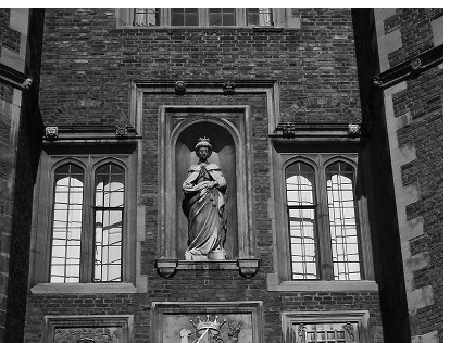}}
  \hspace{0.05in}
    \subfigure[remote sensing]{
    \label{fig:subfig:e}
    \includegraphics[height=2cm]{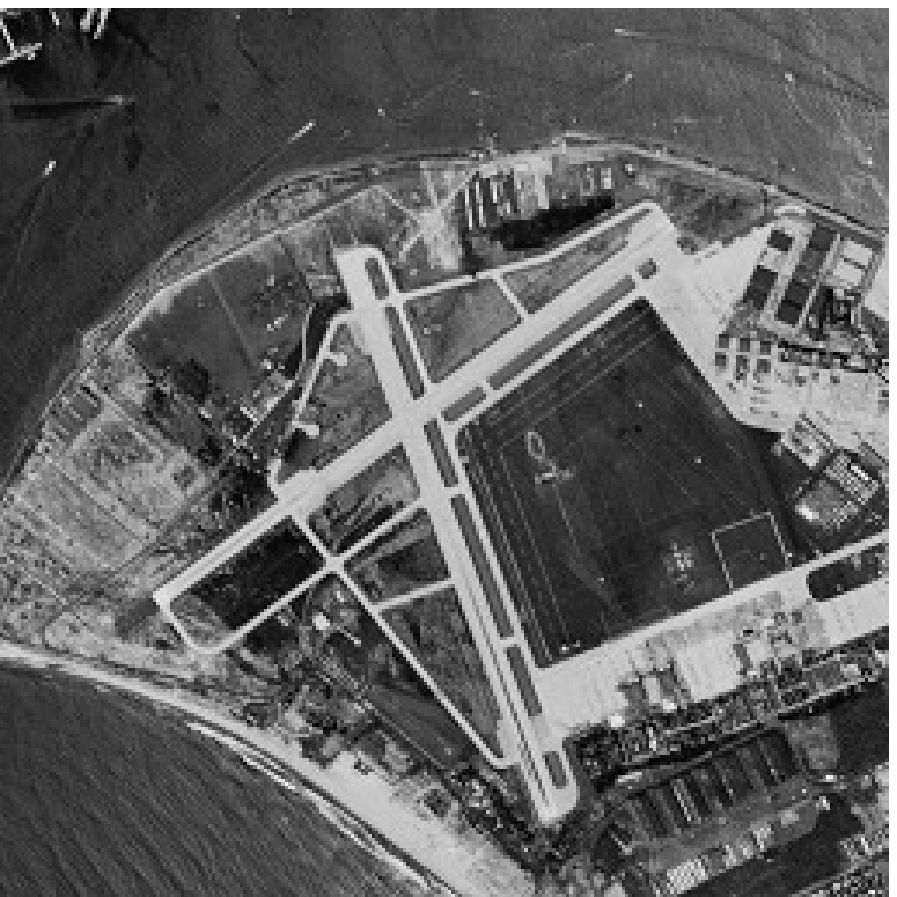}}
  \hspace{0.05in}
  \subfigure[texture]{
    \label{fig:subfig:f}
    \includegraphics[height=2cm]{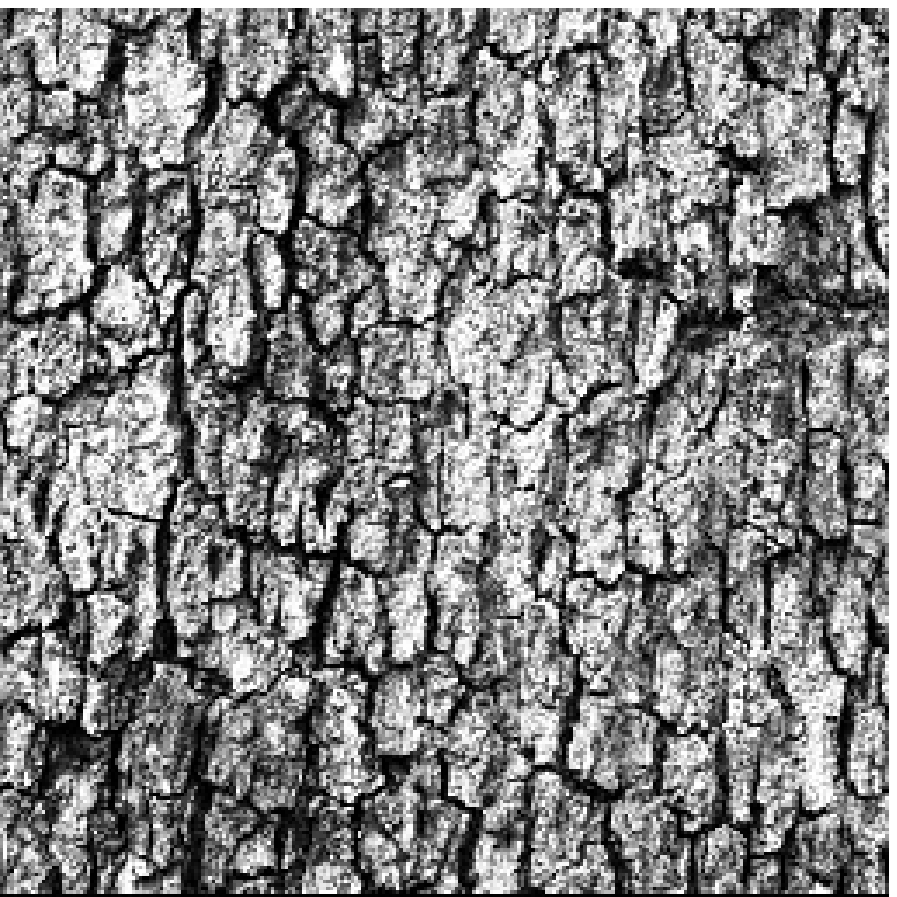}}
  \hspace{0.05in}
  \subfigure[natural beauty]{
    \label{fig:subfig:g}
    \includegraphics[height=2cm]{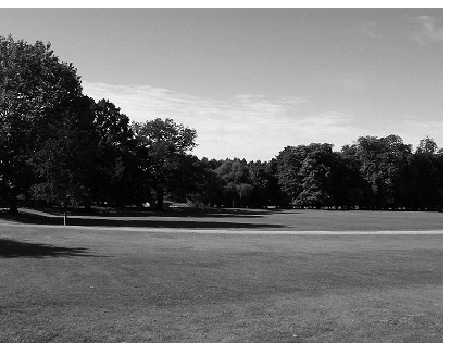}}
  \hspace{0.05in}
  \subfigure[man-made pattern]{
    \label{fig:subfig:h}
    \includegraphics[height=2cm]{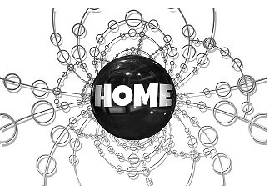}}
  \caption{Samples from the image database (the 800 images) which contains a large range of image types.}
  \label{fig:10}
\end{figure}

We calculate each image's compression ratio $r_{J}$ and show it in
Fig. 11. The maximum value is 0.2364, the minimum value is 0.0148,
the average value is 0.0934, and the variance is only 0.0014. Hence,
in the following, we set
\begin{equation}
r_{J}=0.1
\end{equation}
and the complexity of Step 2 is
\begin{equation}
C_{2}=0.1\cdot \frac{q}{2}2^{2n}=\frac{q}{20}2^{2n}
\end{equation}

\begin{figure}[h]
  \centering
  \includegraphics[width=10cm,keepaspectratio]{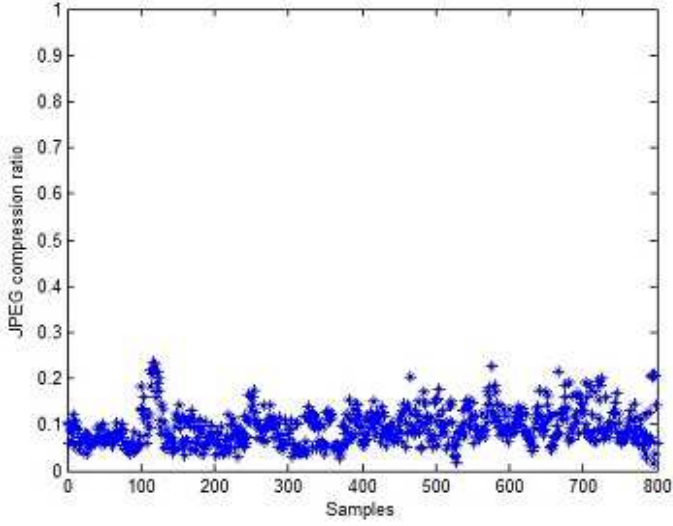}
  \caption{The statistical JPEG compression ratio of the 800 images.}
  \label{fig:11}
\end{figure}

Step 3 is used to store the quantization matrix. Since the
quantization matrix is fixed as Eq. (5) shows, the complexity of
Step 3 (The number of gates used in Step 3) is also fixed. It is
equal to the number of bit 1 in all the quantization values. Hence,
\begin{equation}
C_{3}=208
\end{equation}

The complexity of Step 4 (The number of gates used in Step 4) is
also fixed.
\begin{equation}
\begin{split}
C_{4}=&\text{6 CNOT gates}+\text{1 (6)-CNOT gate}+\text{1
MULERs}+\text{1 (2)-CNOT gate}\\
=&6+1+(q-1)^{2}+4(q-1)+4+\left(4(q-1)-8+2\log_{2}(q-1)\right)\log_{2}(q-1)\\
 &+1\\
=&q^{2}+2q+9+\left(4q-12+2\log_{2}(q-1)\right)\log_{2}(q-1)
\end{split}
\end{equation}

The complexity of Step 5 (The number of gates used in Step 5) is
\begin{equation}
\begin{split}
C_{5}=&\text{The number of bit 1 in all }C_{ij}(u,v)+\text{64 MULERs}\\
      &+\text{128 (8)-CNOT gates}+\text{128 ADDERs}\\
=&\frac{q+4}{2}64+64\left[(q+3)^{2}+4(q+3)-4\right.\\
 &\left.+\left(4(q+3)-8+2\log_{2}(q+3)\right)\log_{2}(q+3)\right]+128+128(8(2q+6)-2)\\
=&64q^{2}+2720q+7232+64(4q+4+2\log_{2}(q+3))\log_{2}(q+3)
\end{split}
\end{equation}

Note that the first item in the third line in Eq. (44) is
$\frac{q+4}{2}64$. That is because: 1)
$C_{ij}(0,0),\cdots,C_{ij}(7,7)$ have 64 components; 2) each
$C_{ij}(u,v)$ is composed by $q+4$ qubits; and 3) each qubit has
equal probability to be 0 and 1.

Hence, the complexity of compressed preparation is
\begin{equation}
\begin{split}
 &C_{2}+C_{3}+C_{4}+C_{5}\\
=&\frac{2^{2n}}{20}q+65q^{2}+2722q+7449+(4q-12+2\log_{2}(q-1))\log_{2}(q-1)\\
 &+64(4q+4+2\log_{2}(q+3))\log_{2}(q+3)
\end{split}
\end{equation}

If our scheme compresses the image preparation,
$$
C_{2}+C_{3}+C_{4}+C_{5}<C
$$
i.e.,
\begin{equation*}
\begin{split}
&\frac{2^{2n}}{20}q+65q^{2}+2722q+7449+(4q-12+2\log_{2}(q-1))\log_{2}(q-1)\\
 &+64(4q+4+2\log_{2}(q+3))\log_{2}(q+3)<\frac{q}{2}2^{2n}
\end{split}
\end{equation*}
Hence, when
\begin{equation}
n>\frac{1}{2}\log_{2}\frac{65q^{2}+2722q+7449+A+B}{0.45q},
\end{equation}
where $A=(4q-12+2\log_{2}(q-1))\log_{2}(q-1)$ and
$B=64(4q+4+2\log_{2}(q+3))\log_{2}(q+3)$, our scheme can compress
the quantum image preparation.

Assume that
$$
m=\frac{1}{2}\log_{2}\frac{65q^{2}+2722q+7449+A+B}{0.45q}
$$
is a function of $q$. The function image is shown in Fig. 12, in
which $q$ is ranged from 4 to 40 to cover almost all the color depth
used in JPEG compression.
\begin{figure}[h]
  \centering
  \includegraphics[width=10cm,keepaspectratio]{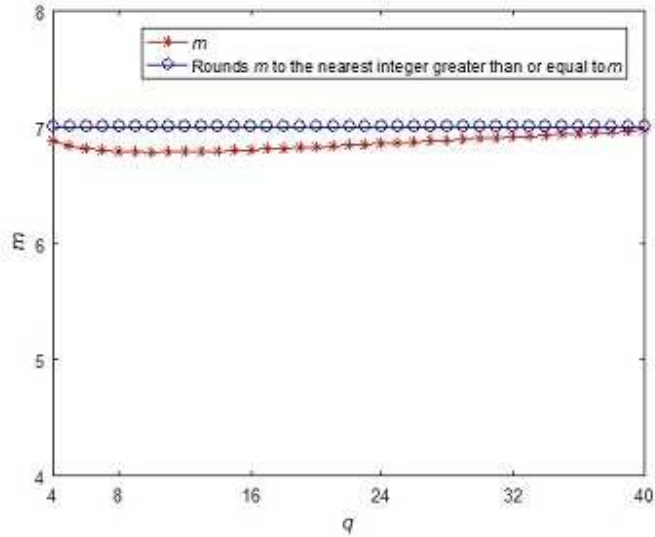}
  \caption{The relationship between the color depth $q$ and the $m$. Because $n$ is a positive integer,
  we also round $m$ to the nearest integer greater than or equal to $m$. That is to say, the line with small circles gives the lowest value of $n$.}
  \label{fig:12}
\end{figure}

Fig. 12 tells us that as long as the image size is bigger than
$2^{7}\times2^{7}=128\times128$, our scheme can compress the image.
It is a easy to achieve requirement because most of the common image
sizes are bigger than $128\times128$.

\begin{definition}
Quantum image compression ratio
$$
r=1-\frac{\text{The number of operations after
compression}}{\text{The number of operations before compression}}
$$
\end{definition}

Due to
$$
r=1-\frac{C_{2}+C_{3}+C_{4}+C_{5}}{C},
$$
the compression ratio of our scheme is
\begin{equation}
r=1-\frac{\frac{2^{2n}}{20}q+65q^{2}+2722q+7449+A+B}{\frac{q}{2}2^{2n}}
\end{equation}

Fig. 13 gives the $r$ changed with $n$ and $q$. It
tells us that:
\begin{enumerate}[(1)]
\item No matter how much the value of $q$, when $n\geq10$, i.e., the
image size is equal to or bigger than $1024\times1024$, $r$ is
approximately 0.9. That is to say, the JPEG complexity is only
$\frac{1}{10}$ of the original complexity.
\item The bigger the $n$, the higher the compression ratio $r$.
\end{enumerate}

The reason to these two items is that
$$
r=1-\frac{C_{2}+C_{3}+C_{4}+C_{5}}{C}=0.9-\frac{C_{3}+C_{4}+C_{5}}{\frac{q}{2}2^{2n}}
$$
and $C_{3}$, $C_{4}$ and $C_{5}$ are independent of $n$, hence as
$n$ increases, $r$ increases rapidly and the limit is 0.9.

\begin{figure}[h]
  \centering
  \includegraphics[width=10cm,keepaspectratio]{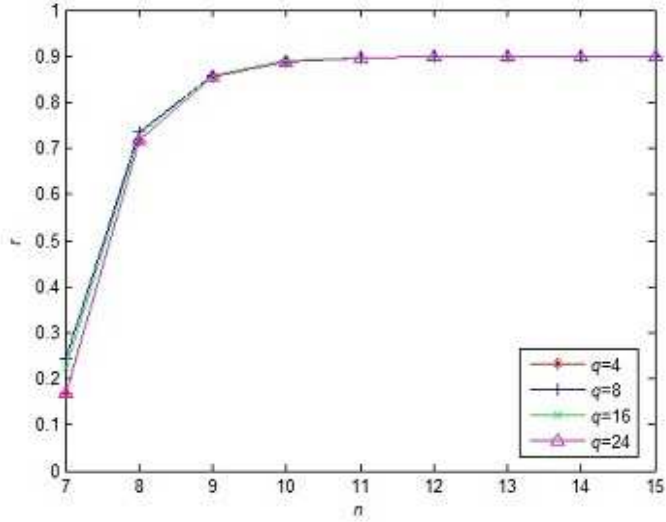}
  \caption{The quantum image compression ratio $r$ is changed with the image size $n$ and color depth $q$.}
  \label{fig:13}
\end{figure}

We also compare the compression ratio of our scheme with BEC in
Table 1. However, as stated in Section 2.1, if the preprocessing is
taken into account, BEC's complexity is too high to bear. Hence, we
only test 3 $256\times256$ images with $q=8$. It can be seen that
our scheme's compression ratio is higher than that of BEC scheme.

\begin{table}
  \centering
  \caption{The compression ratio (not include preprocessing).}
  \begin{tabular}{|c|c|c|c|c|}
  \hline
  \multirow{2}{*}{Image} & \multicolumn{2}{c|}{Compression ratio} & \multicolumn{2}{c|}{\tabincell{c}{The number of operations \\after/before compression (GQIR)}}\\
  \cline{2-5} & \tabincell{c}{our scheme\\(JPEG)} & \tabincell{c}{BEC} & \tabincell{c}{our scheme\\(JPEG)} & \tabincell{c}{BEC}\\
  \hline
  \parbox[c]{0.72in}{\includegraphics[scale=0.08]{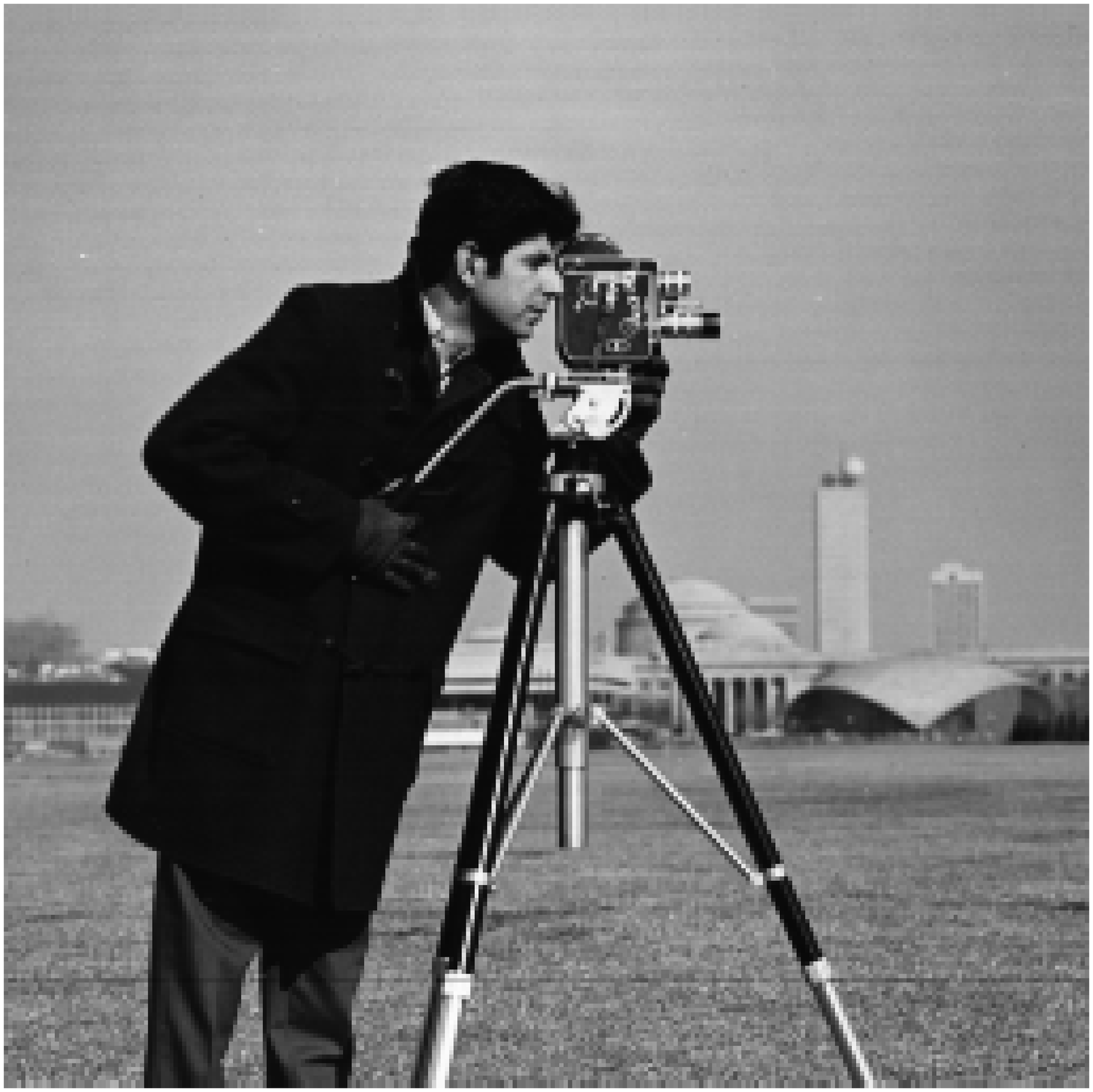}\\Cameraman} & 84.66\% & 69.07\% & 37638/245366 & 75901/245366\\
  \hline
  \parbox[c]{0.72in}{\includegraphics[scale=0.08]{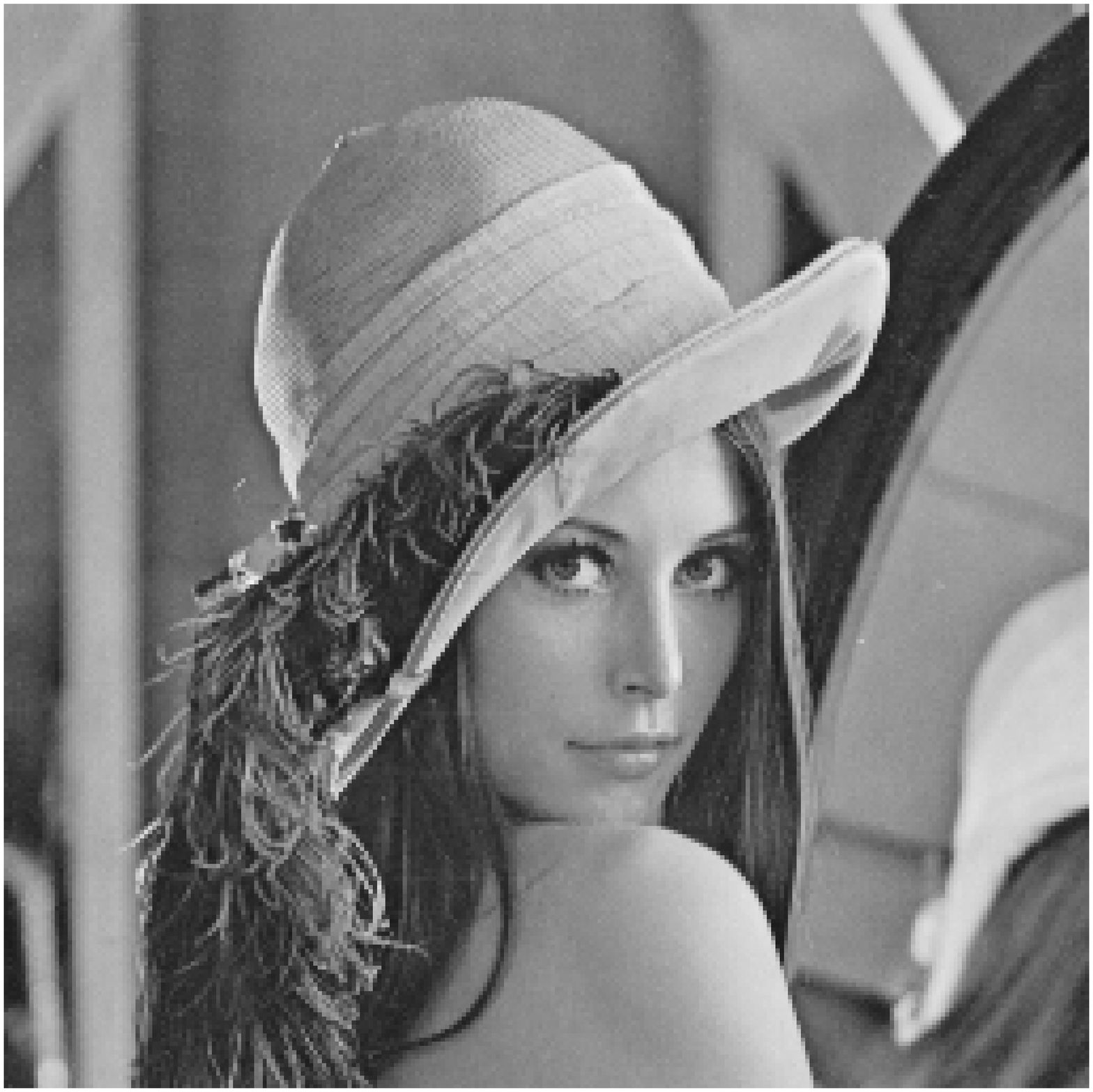}\\Lena} & 85.09\% & 66.17\% & 38450/257833 & 87217/257833\\
  \hline
  \parbox[c]{0.72in}{\includegraphics[scale=0.08]{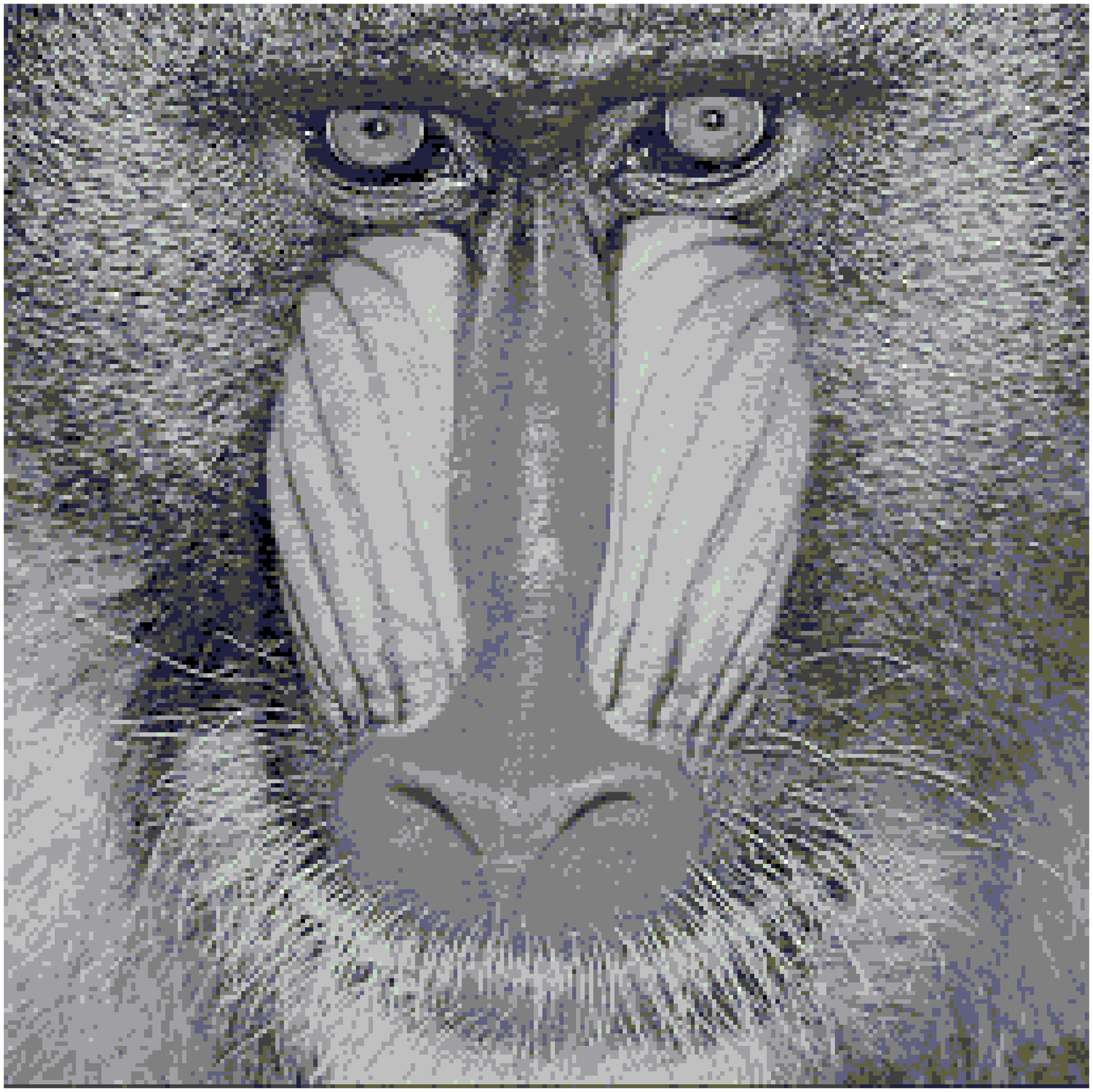}\\Baboon} & 67.58\% & 66.00\% & 89911/277344 & 94307/277344\\
  \hline
  \end{tabular}
\end{table}

\subsection{Classical part}

Our scheme and BEC have a preprocessing step done in classical
computers.

From Eq. (4) and (6), it is obvious that the complexity of DCT and
quantization are both $O(2^{2n})$. Hence, the preprocessing
complexity of our scheme is $O(2\cdot2^{2n})$. From Eq. (3), the
preprocessing complexity of BEC is $O(2n\cdot q\cdot2^{4n})$.
$$
O(2n\cdot q\cdot2^{4n})\gg O(2\cdot2^{2n})
$$

Furthermore, there are fast algorithms for DCT which can improve the
efficiency [24]. Fast DCT has been widely used in classical image
compression. When you use your phone to take a picture, you do DCT
one time because the picture has been compressed and its filename
extension is ``jpg''. We do not plane to describe fast DCT
theoretically as the length limit of the paper. However, some data
are given to compare our scheme and BEC intuitively (see Table 2).
The experiment environment is a desktop with Intel(R) Core(TM)
i5-4590 CPU 3.3GHz, 8GB Ram equipped with MATLAB R2016b.

\begin{table}
  \centering
  \caption{The preprocessing time ($q=8$).}
  \begin{tabular}{|c|c|c|c|c|}
  \hline
  image name & $n$ & image size & our scheme & BEC\\
  \hline
  Cameraman & 8 & $256\times256$ & 0.176 second & ---\\
  \hline
  Lena & 8 & $256\times256$ & 0.164 second & 5.54 hour\\
  \hline
  Baboon & 8 & $256\times256$ & 0.164 second & ---\\
  \hline
  Cameraman & 9 & $512\times512$ & 0.516 second & ---\\
  \hline
  Lena & 9 & $512\times512$ & 0.505 second & more than 3 days\\
  \hline
  Baboon & 9 & $512\times512$ & 0.505 second & ---\\
  \hline
  \end{tabular}
\\Note: 1) ``---'' indicates that we do not do that experiment. 2)
``more than 3 days'' because the experiment has been done for 3 days
but has not been finished, so we stopped it.
\end{table}

It is obviously that our scheme is faster than BEC.

\section{Visual effects}

Unlike GQIR preparation and BEC, our scheme is a lossy method which
may affect the quality of the image. However, JPEG can help us to
ensure the visual quality of the images. In this section, some
examples are given to display the visual effects.

Due to the condition that the physical quantum hardware is not
affordable for us to execute our protocol, we just make the
simulations on the classical computer mentioned in Section 4.2.

Fig. 14 gives the visual effects of our scheme.
\begin{figure}[h]
  \centering
  \subfigure[uncompressed cameraman]{
    \label{fig:subfig:a}
    \includegraphics[height=5cm]{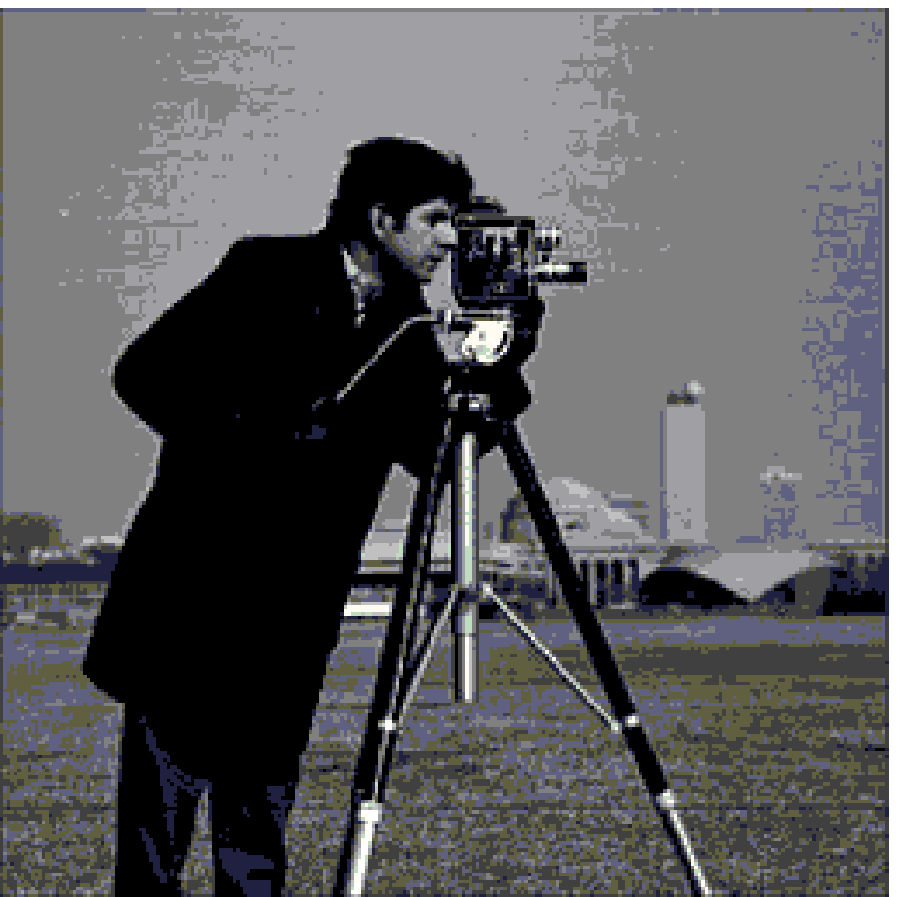}}
  \hspace{0.05in}
  \subfigure[compressed cameraman]{
    \label{fig:subfig:b}
    \includegraphics[height=5cm]{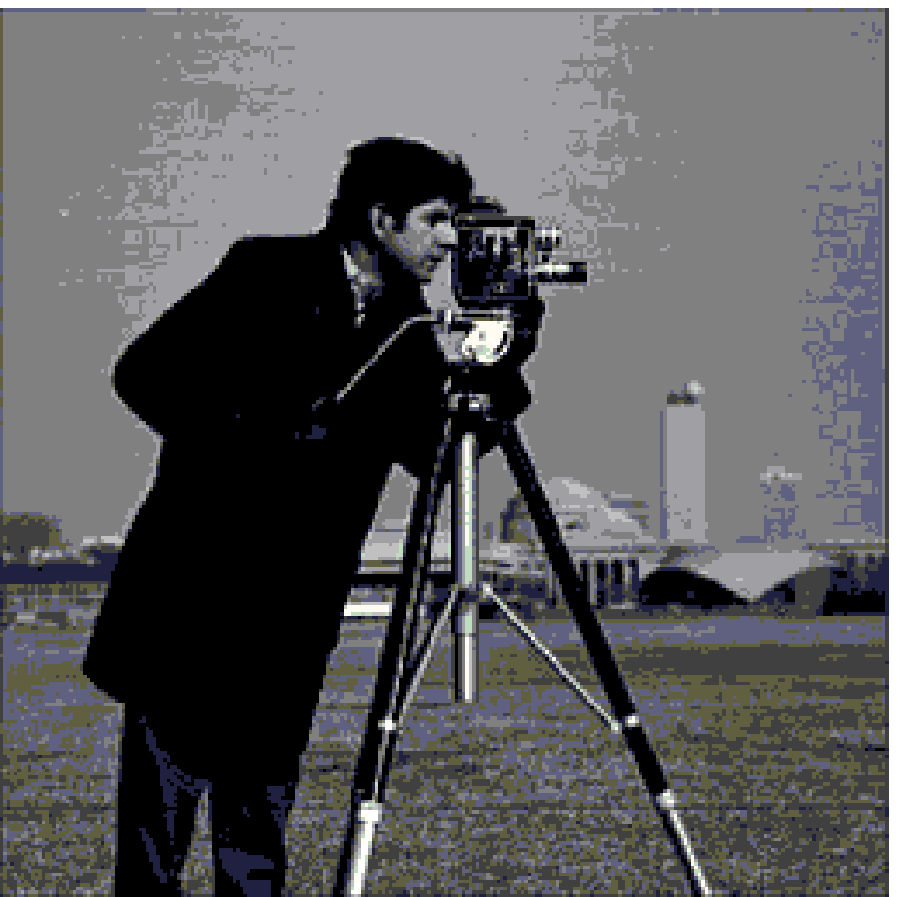}}
  \hspace{0.05in}
  \subfigure[uncompressed Lena]{
    \label{fig:subfig:c}
    \includegraphics[height=5cm]{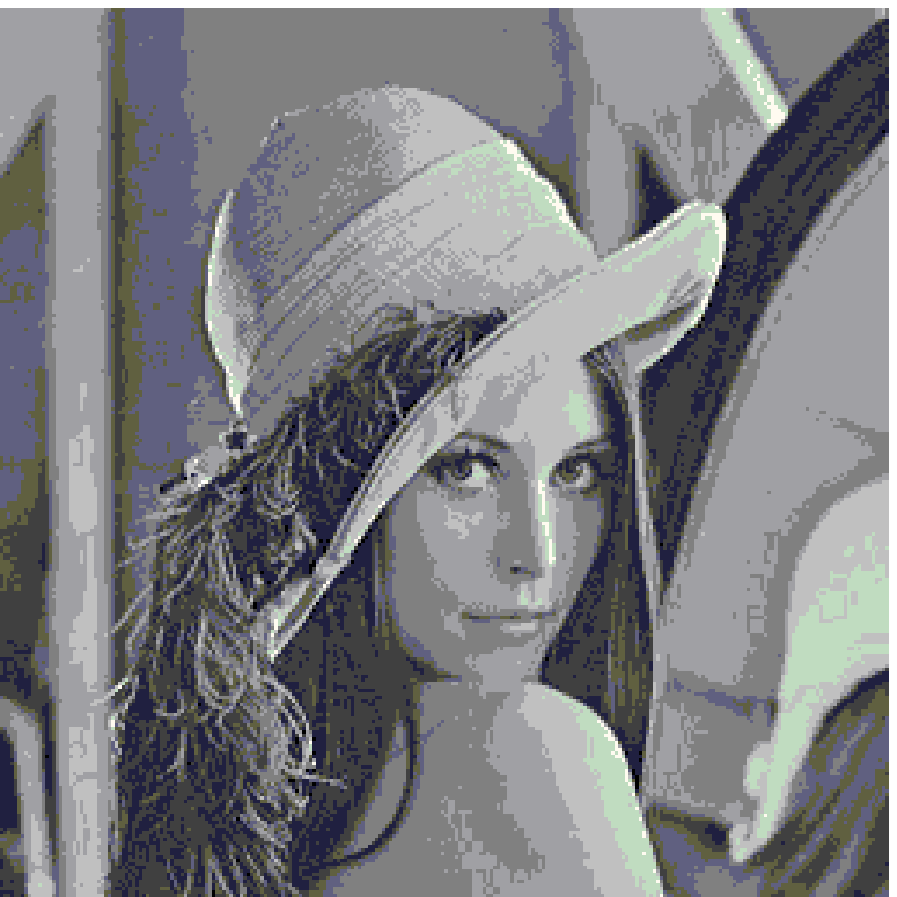}}
  \hspace{0.05in}
  \subfigure[compressed Lena]{
    \label{fig:subfig:d}
    \includegraphics[height=5cm]{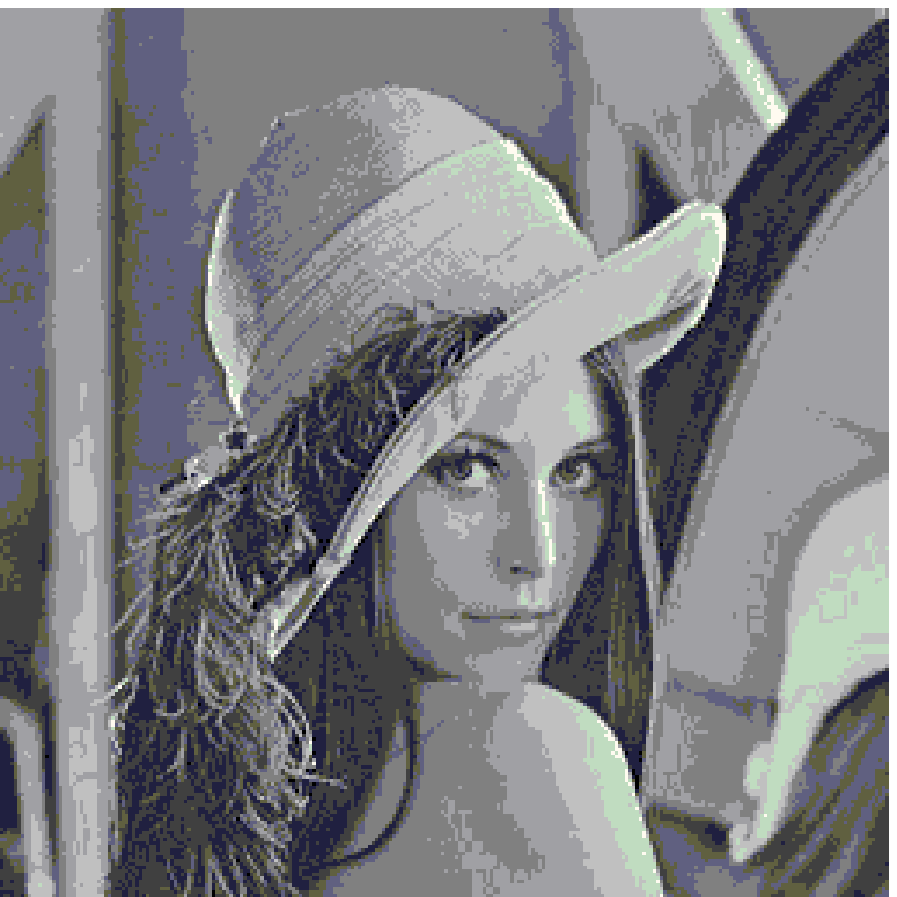}}
  \hspace{0.05in}
    \subfigure[uncompressed baboon]{
    \label{fig:subfig:e}
    \includegraphics[height=5cm]{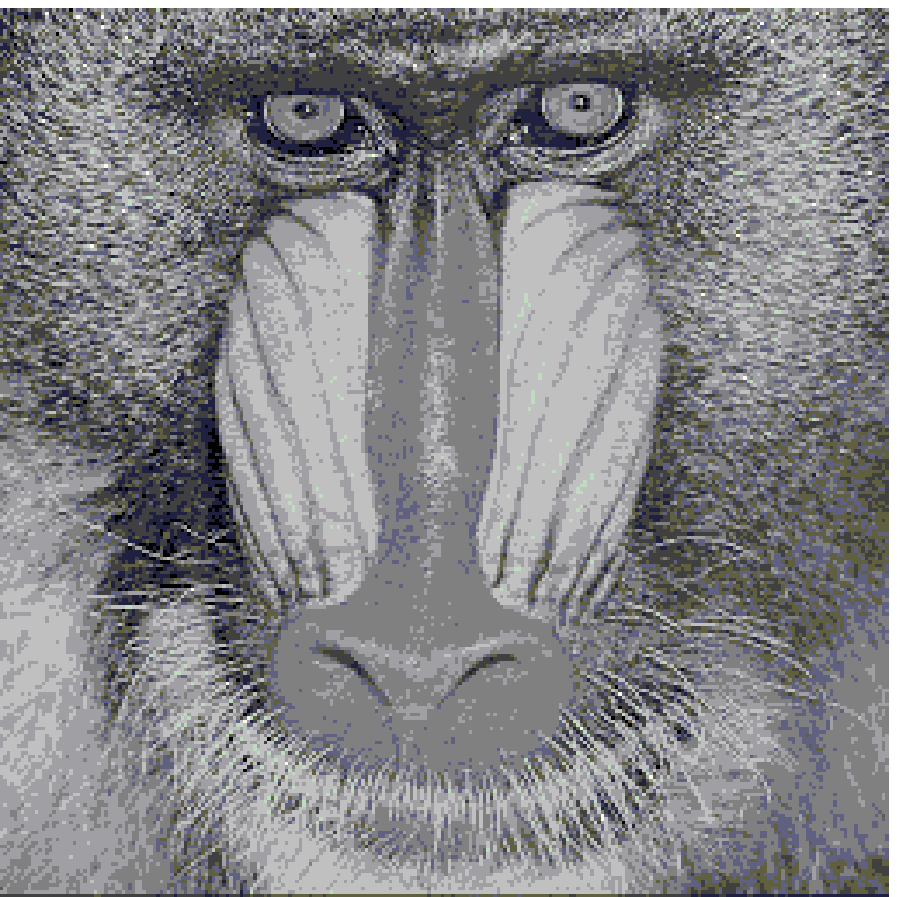}}
  \hspace{0.05in}
  \subfigure[compressed baboon]{
    \label{fig:subfig:f}
    \includegraphics[height=5cm]{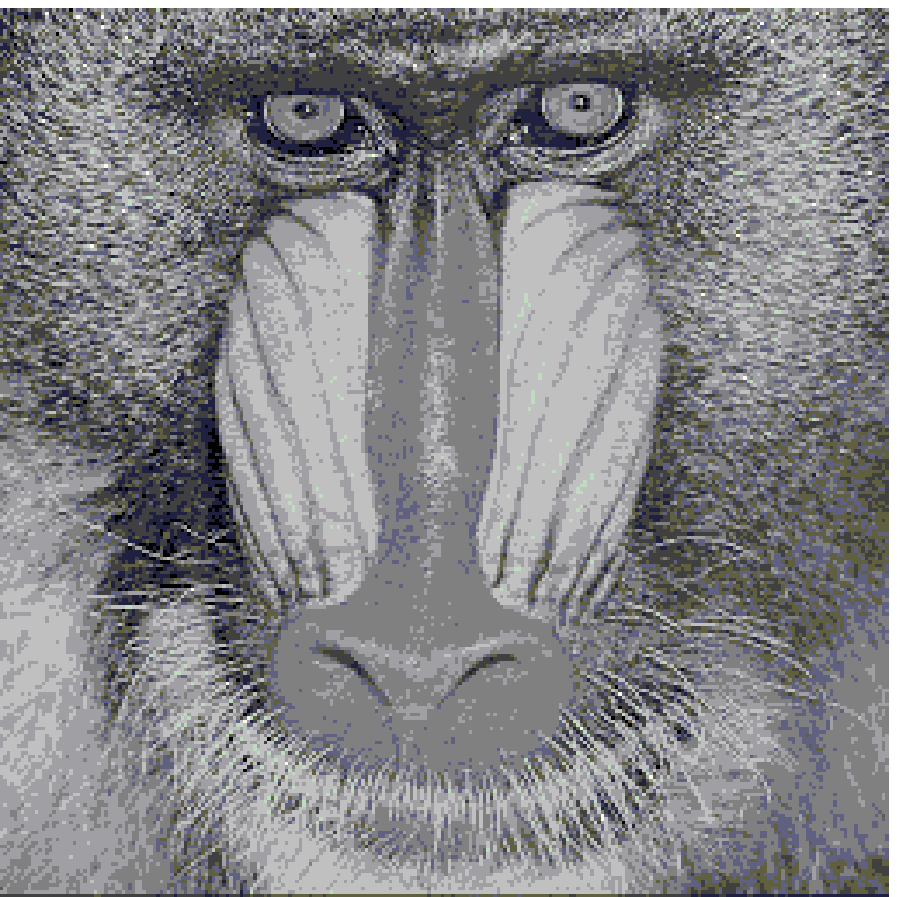}}
  \caption{Visual effects of our scheme. Images in the left are uncompressed and images in the right are compressed.}
  \label{fig:14}
\end{figure}

The peak-signal-to-noise ratio (PSNR), being one of the most used
quantity in classical for comparing the fidelity of two images, will
be used to evaluate the quality of the compressed images. By
assuming $I$ and $I'$ are two $2^{n}\times2^n$ images, the PSNR is
defined by
$$
\text{PSNR}=20\log_{10}\left(\frac{MAX_{I}}{\sqrt{\frac{1}{2^{4n}}\sum_{i=0}^{2^{n}-1}\sum_{j=0}^{2^{n}-1}[I(i,j)-I'(i,j)]}}\right)
$$
where, $MAX_{I}$ is the maximum possible pixel value of the image.
Table 3 gives the PSNR between images before and after compression.
\begin{table}
  \centering
  \caption{The PSNR values between images before and after compression.}
  \begin{tabular}{|c|c|}
  \hline
  image & PSNR\\
  \hline
  Cameraman & 38.0683\\
  \hline
  Lena & 35.7164\\
  \hline
  Baboon & 27.9511\\
  \hline
  \end{tabular}
\end{table}

From Fig. 14 and Table 3, we can see that our scheme does not affect
the images' visual effect, and the PSNR is acceptable.

\section{Conclusion}

This paper focuses on quantum image preparation and gives a new
quantum image compression scheme based on JPEG to prepare a GQIR
image. Compared with BEC, our scheme has the following advantages:
\begin{enumerate}
  \item Its preprocessing (DCT and quantization) is simple and fast.
  \item As long as Eq. (46) is satisfied, its compression ratio is higher than
  that of BEC. Moreover, Eq. (46) is a loose condition because it
  requires that the size of the image is only bigger than or equal
  to $128\times128$.
\end{enumerate}

\begin{acknowledgements}
The authors thank Prof. Sabre Kais at Purdue University for his
valuable suggestions.
\end{acknowledgements}



\end{document}